%\input harvmac

%%%%%%%%%%%%%%%%%%  tex macros for preprints, cm version %%%%%%%%%%%%%%
%         (P. Ginsparg <ginsparg@lanl.gov>, last updated 7/94)
%                if confused, type `b' in response to query
%           hypertex extensions (still provisional), 7/26/94
%
%---------------------------------------------------------------------%
%\input hyperbasics %comment out this line to restore non-hyper functionality
%
%% site dependent options:
%% \unredoffs and \redoffs define horizontal and vertical offsets
%% respectively for unreduced and reduced modes. \speclscape defines
%% the \special{} call that sets printer to landscape (sideways) mode.
%% from standard set below, leave uncommented as appropriate or redefine
%
%%% next 400dpi
\def\unredoffs{} \def\redoffs{\voffset=-.31truein\hoffset=-.48truein}
\def\speclscape{}
%\def\speclscape{\special{papersize=11in,8.5in}}
%
%%% apple lw
%\def\unredoffs{} \def\redoffs{\voffset=-.31truein\hoffset=-.59truein}
%\def\speclscape{\special{ps: landscape}}
%
%%% qms lasergrafix:
%\def\unredoffs{} \def\redoffs{\voffset=-.4truein\hoffset=.125truein}
%\def\speclscape{\special{qms: landscape}}
%
%%% saclay A4 paper:
%\def\unredoffs{\hoffset-.14truein\voffset-.2truein}
%\def\redoffs{\voffset=-.45truein\hoffset=-.21truein}
%\def\speclscape{\special{landscape}}
%
%---------------------------------------------------------------------%
%
\newbox\leftpage \newdimen\fullhsize \newdimen\hstitle \newdimen\hsbody
\tolerance=1000\hfuzz=2pt
\catcode`\@=11 % This allows us to modify PLAIN macros.
\ifx\hyperdef\UNd@FiNeD\def\hyperdef#1#2#3#4{#4}\def\hyperref#1#2#3#4{#4}\fi
\def\bigans{b }
\def\answ{b }
%\message{ big or little (b/l)? }\read-1 to\answ
%
\ifx\answ\bigans\message{(This will come out unreduced.}
\magnification=1200\unredoffs\baselineskip=16pt plus 2pt minus 1pt
\hsbody=\hsize \hstitle=\hsize %take default values for unreduced format
\else\message{(This will be reduced.} \let\l@r=L
\magnification=1000\baselineskip=16pt plus 2pt minus 1pt \vsize=7truein
\redoffs \hstitle=8truein\hsbody=4.75truein\fullhsize=10truein\hsize=\hsbody
\output={\ifnum\pageno=0 %%% This is the HUTP version
  \shipout\vbox{\speclscape{\hsize\fullhsize\makeheadline}
    \hbox to \fullhsize{\hfill\pagebody\hfill}}\advancepageno
  \else
  \almostshipout{\leftline{\vbox{\pagebody\makefootline}}}\advancepageno
  \fi}
\def\almostshipout#1{\if L\l@r \count1=1 \message{[\the\count0.\the\count1]}
      \global\setbox\leftpage=#1 \global\let\l@r=R
 \else \count1=2
  \shipout\vbox{\speclscape{\hsize\fullhsize\makeheadline}
      \hbox to\fullhsize{\box\leftpage\hfil#1}}  \global\let\l@r=L\fi}
\fi
%---------------------------------------------------------------------
%
\newcount\yearltd\yearltd=\year\advance\yearltd by -2000

\def\Title#1#2{\nopagenumbers\abstractfont\hsize=\hstitle\rightline{#1}%
\vskip 1in\centerline{\titlefont #2}\abstractfont\vskip .5in\pageno=0}
\def\Date#1{\vfill\leftline{#1}\tenpoint\supereject\global\hsize=\hsbody%
\footline={\hss\tenrm\hyperdef\hypernoname{page}\folio\folio\hss}}%
% (restores pagenumbers)
%
%       use following instead of \Date on the preliminary draft,
%       puts date/time on each page in big mode, writes labels in margins

\def\draftmode{\message{ DRAFTMODE }\def\draftdate{{\rm preliminary draft:
\number\month/\number\day/\number\yearltd\ \ \hourmin}}%
\headline={\hfil\draftdate}\writelabels\baselineskip=20pt plus 2pt minus 2pt
 {\count255=\time\divide\count255 by 60 \xdef\hourmin{\number\count255}
  \multiply\count255 by-60\advance\count255 by\time
  \xdef\hourmin{\hourmin:\ifnum\count255<10 0\fi\the\count255}}}
%       use \nolabels to get rid of eqn, ref, and fig labels in draft mode
\def\nolabels{\def\wrlabeL##1{}\def\eqlabeL##1{}\def\reflabeL##1{}}
\def\writelabels{\def\wrlabeL##1{\leavevmode\vadjust{\rlap{\smash%
{\line{{\escapechar=` \hfill\rlap{\sevenrm\hskip.03in\string##1}}}}}}}%
\def\eqlabeL##1{{\escapechar-1\rlap{\sevenrm\hskip.05in\string##1}}}%
\def\reflabeL##1{\noexpand\llap{\noexpand\sevenrm\string\string\string##1}}}
\nolabels
%
% tagged sec numbers
\global\newcount\secno \global\secno=0
\global\newcount\meqno \global\meqno=1
\def\s@csym{}
\def\newsec#1{\global\advance\secno by1%
{\toks0{#1}\message{(\the\secno. \the\toks0)}}%
%\ifx\answ\bigans \vfill\eject \else \bigbreak\bigskip \fi  %if desired
\global\subsecno=0\eqnres@t\let\s@csym\secsym\xdef\secn@m{\the\secno}\noindent
{\bf\hyperdef\hypernoname{section}{\the\secno}{\the\secno.} #1}%
\writetoca{{\string\hyperref{}{section}{\the\secno}{\the\secno.}} {#1}}%
\par\nobreak\medskip\nobreak}
\def\eqnres@t{\xdef\secsym{\the\secno.}\global\meqno=1\bigbreak\bigskip}
\def\sequentialequations{\def\eqnres@t{\bigbreak}}\xdef\secsym{}
\global\newcount\subsecno \global\subsecno=0
\def\subsec#1{\global\advance\subsecno by1%
{\toks0{#1}\message{(\s@csym\the\subsecno. \the\toks0)}}%
\ifnum\lastpenalty>9000\else\bigbreak\fi
\noindent{\it\hyperdef\hypernoname{subsection}{\secn@m.\the\subsecno}%
{\secn@m.\the\subsecno.} #1}\writetoca{\string\quad
{\string\hyperref{}{subsection}{\secn@m.\the\subsecno}{\secn@m.\the\subsecno.}}
{#1}}\par\nobreak\medskip\nobreak}
\def\appendix#1#2{\global\meqno=1\global\subsecno=0\xdef\secsym{\hbox{#1.}}%
\bigbreak\bigskip\noindent{\bf Appendix \hyperdef\hypernoname{appendix}{#1}%
{#1.} #2}{\toks0{(#1. #2)}\message{\the\toks0}}%
\xdef\s@csym{#1.}\xdef\secn@m{#1}%
\writetoca{\string\hyperref{}{appendix}{#1}{Appendix {#1.}} {#2}}%
\par\nobreak\medskip\nobreak}
%
%       \eqn\label{a+b=c}	gives displayed equation, numbered
%				consecutively within sections.
%     \eqnn and \eqna define labels in advance (of eqalign?)
%
\def\checkm@de#1#2{\ifmmode{\def\f@rst##1{##1}\hyperdef\hypernoname{equation}%
{#1}{#2}}\else\hyperref{}{equation}{#1}{#2}\fi}
\def\eqnn#1{\DefWarn#1\xdef #1{(\noexpand\relax\noexpand\checkm@de%
{\s@csym\the\meqno}{\secsym\the\meqno})}%
\wrlabeL#1\writedef{#1\leftbracket#1}\global\advance\meqno by1}
\def\f@rst#1{\c@t#1a\em@ark}\def\c@t#1#2\em@ark{#1}
\def\eqna#1{\DefWarn#1\wrlabeL{#1$\{\}$}%
\xdef #1##1{(\noexpand\relax\noexpand\checkm@de%
{\s@csym\the\meqno\noexpand\f@rst{##1}}{\hbox{$\secsym\the\meqno##1$}})}
\writedef{#1\numbersign1\leftbracket#1{\numbersign1}}\global\advance\meqno by1}
\def\eqn#1#2{\DefWarn#1%
\xdef #1{(\noexpand\hyperref{}{equation}{\s@csym\the\meqno}%
{\secsym\the\meqno})}$$#2\eqno(\hyperdef\hypernoname{equation}%
{\s@csym\the\meqno}{\secsym\the\meqno})\eqlabeL#1$$%
\writedef{#1\leftbracket#1}\global\advance\meqno by1}
\def\xeqn{\expandafter\xe@n}\def\xe@n(#1){#1}
\def\xeqna#1{\expandafter\xe@n#1}
\def\eqns#1{(\e@ns #1{\hbox{}})}
\def\e@ns#1{\ifx\UNd@FiNeD#1\message{eqnlabel \string#1 is undefined.}%
\xdef#1{(?.?)}\fi{\let\hyperref=\relax\xdef\next{#1}}%
\ifx\next\em@rk\def\next{}\else%
\ifx\next#1\xeqn#1\else\def\n@xt{#1}\ifx\n@xt\next#1\else\xeqna#1\fi
\fi\let\next=\e@ns\fi\next}

\def\DefWarn#1{\ifx\UNd@FiNeD#1\else
\immediate\write16{*** WARNING: the label \string#1 is already defined ***}\fi}
%
%			 footnotes
\newskip\footskip\footskip14pt plus 1pt minus 1pt %sets footnote baselineskip
\def\footnotefont{\ninepoint}\def\f@t#1{\footnotefont #1\@foot}
\def\f@@t{\baselineskip\footskip\bgroup\footnotefont\aftergroup\@foot\let\next}
\setbox\strutbox=\hbox{\vrule height9.5pt depth4.5pt width0pt}
\global\newcount\ftno \global\ftno=0
\def\foot{\global\advance\ftno by1\def\foot@rg{\hyperref{}{footnote}%
{\the\ftno}{\the\ftno}\xdef\foot@rg{\noexpand\hyperdef\noexpand\hypernoname%
{footnote}{\the\ftno}{\the\ftno}}}\footnote{$^{\foot@rg}$}}
%
%say \footend to put footnotes at end
%will cause problems if \ref used inside \foot, instead use \nref before
\newwrite\ftfile
\def\footend{\def\foot{\global\advance\ftno by1\chardef\wfile=\ftfile
%%$^{\the\ftno}$\ifnum\ftno=1\immediate\openout\ftfile=\jobname.fts\fi%
\hyperref{}{footnote}{\the\ftno}{$^{\the\ftno}$}%
\ifnum\ftno=1\immediate\openout\ftfile=\jobname.fts\fi%
\immediate\write\ftfile{\noexpand\smallskip%
%%\noexpand\item{f\the\ftno:\ }\pctsign}\findarg}%
\noexpand\item{\noexpand\hyperdef\noexpand\hypernoname{footnote}
{\the\ftno}{f\the\ftno}:\ }\pctsign}\findarg}%
\def\footatend{\vfill\eject\immediate\closeout\ftfile{\parindent=20pt
\centerline{\bf Footnotes}\nobreak\bigskip\input \jobname.fts }}}
\def\footatend{}
%
%     \ref\label{text}
% generates a number, assigns it to \label, generates an entry.
% To list the refs on a separate page,  \listrefs
%
\global\newcount\refno \global\refno=1
\newwrite\rfile
\def\ref{[\hyperref{}{reference}{\the\refno}{\the\refno}]\nref}
\def\nref#1{\DefWarn#1%
\xdef#1{[\noexpand\hyperref{}{reference}{\the\refno}{\the\refno}]}%
\writedef{#1\leftbracket#1}%
\ifnum\refno=1\immediate\openout\rfile=\jobname.refs\fi
\chardef\wfile=\rfile\immediate\write\rfile{\noexpand\item{[\noexpand\hyperdef%
\noexpand\hypernoname{reference}{\the\refno}{\the\refno}]\ }%
\reflabeL{#1\hskip.31in}\pctsign}\global\advance\refno by1\findarg}
%	horrible hack to sidestep tex \write limitation
\def\findarg#1#{\begingroup\obeylines\newlinechar=`\^^M\pass@rg}
{\obeylines\gdef\pass@rg#1{\writ@line\relax #1^^M\hbox{}^^M}%
\gdef\writ@line#1^^M{\expandafter\toks0\expandafter{\striprel@x #1}%
\edef\next{\the\toks0}\ifx\next\em@rk\let\next=\endgroup\else\ifx\next\empty%
\else\immediate\write\wfile{\the\toks0}\fi\let\next=\writ@line\fi\next\relax}}
\def\striprel@x#1{} \def\em@rk{\hbox{}}
\def\lref{\begingroup\obeylines\lr@f}
\def\lr@f#1#2{\DefWarn#1\gdef#1{\let#1=\UNd@FiNeD\ref#1{#2}}\endgroup\unskip}

\def\addref#1{\immediate\write\rfile{\noexpand\item{}#1}} %now unnecessary
\def\listrefs{\footatend\vfill\supereject\immediate\closeout\rfile\writestoppt
\baselineskip=\footskip\centerline{{\bf References}}\bigskip{\parindent=20pt%
\frenchspacing\escapechar=` \input \jobname.refs\vfill\eject}\nonfrenchspacing}
\def\startrefs#1{\immediate\openout\rfile=\jobname.refs\refno=#1}
\def\xref{\expandafter\xr@f}\def\xr@f[#1]{#1}
\def\refs#1{\count255=1[\r@fs #1{\hbox{}}]}
\def\r@fs#1{\ifx\UNd@FiNeD#1\message{reflabel \string#1 is undefined.}%
\nref#1{need to supply reference \string#1.}\fi%
\vphantom{\hphantom{#1}}{\let\hyperref=\relax\xdef\next{#1}}%
\ifx\next\em@rk\def\next{}%
\else\ifx\next#1\ifodd\count255\relax\xref#1\count255=0\fi%
\else#1\count255=1\fi\let\next=\r@fs\fi\next}
%

%
% this is ugly, but moore insists
\newwrite\ffile\global\newcount\figno \global\figno=1
\def\fig{fig.~\hyperref{}{figure}{\the\figno}{\the\figno}\nfig}
\def\nfig#1{\DefWarn#1%
\xdef#1{fig.~\noexpand\hyperref{}{figure}{\the\figno}{\the\figno}}%
\writedef{#1\leftbracket fig.\noexpand~\xfig#1}%
\ifnum\figno=1\immediate\openout\ffile=\jobname.figs\fi\chardef\wfile=\ffile%
{\let\hyperref=\relax
\immediate\write\ffile{\noexpand\medskip\noexpand\item{Fig.\ %
\noexpand\hyperdef\noexpand\hypernoname{figure}{\the\figno}{\the\figno}. }
\reflabeL{#1\hskip.55in}\pctsign}}\global\advance\figno by1\findarg}
\def\listfigs{\vfill\eject\immediate\closeout\ffile{\parindent40pt
\baselineskip14pt\centerline{{\bf Figure Captions}}\nobreak\medskip
\escapechar=` \input \jobname.figs\vfill\eject}}
\def\xfig{\expandafter\xf@g}\def\xf@g fig.\penalty\@M\ {}
\def\figs#1{figs.~\f@gs #1{\hbox{}}}
\def\f@gs#1{{\let\hyperref=\relax\xdef\next{#1}}\ifx\next\em@rk\def\next{}\else
\ifx\next#1\xfig #1\else#1\fi\let\next=\f@gs\fi\next}
\def\figin{\epsfcheck\figin}\def\figins{\epsfcheck\figins}
\def\epsfcheck{\ifx\epsfbox\UNd@FiNeD
\message{(NO epsf.tex, FIGURES WILL BE IGNORED)}
\gdef\figin##1{\vskip2in}\gdef\figins##1{\hskip.5in}% blank space instead
\else\message{(FIGURES WILL BE INCLUDED)}%
\gdef\figin##1{##1}\gdef\figins##1{##1}\fi}
\def\DefWarn#1{}
\def\figinsert{\goodbreak\midinsert}
\def\ifig#1#2#3{\DefWarn#1\xdef#1{fig.~\noexpand\hyperref{}{figure}%
{\the\figno}{\the\figno}}\writedef{#1\leftbracket fig.\noexpand~\xfig#1}%
\figinsert\figin{\centerline{#3}}\medskip\centerline{\vbox{\baselineskip12pt
\advance\hsize by -1truein\noindent\wrlabeL{#1=#1}\footnotefont%
{\bf Fig.~\hyperdef\hypernoname{figure}{\the\figno}{\the\figno}:} #2}}
\bigskip\endinsert\global\advance\figno by1}
\newwrite\lfile
{\escapechar-1\xdef\pctsign{\string\%}\xdef\leftbracket{\string\{}
\xdef\rightbracket{\string\}}\xdef\numbersign{\string\#}}
\def\writedefs{\immediate\openout\lfile=\jobname.defs \def\writedef##1{%
{\let\hyperref=\relax\let\hyperdef=\relax\let\hypernoname=\relax
 \immediate\write\lfile{\string\def\string##1\rightbracket}}}}%
\def\writestop{\def\writestoppt{\immediate\write\lfile{\string\pageno
 \the\pageno\string\startrefs\leftbracket\the\refno\rightbracket
 \string\def\string\secsym\leftbracket\secsym\rightbracket
 \string\secno\the\secno\string\meqno\the\meqno}\immediate\closeout\lfile}}
\def\writestoppt{}\def\writedef#1{}
\def\seclab#1{\DefWarn#1%
\xdef #1{\noexpand\hyperref{}{section}{\the\secno}{\the\secno}}%
\writedef{#1\leftbracket#1}\wrlabeL{#1=#1}}
\def\subseclab#1{\DefWarn#1%
\xdef #1{\noexpand\hyperref{}{subsection}{\secn@m.\the\subsecno}%
{\secn@m.\the\subsecno}}\writedef{#1\leftbracket#1}\wrlabeL{#1=#1}}
\def\applab#1{\DefWarn#1%
\xdef #1{\noexpand\hyperref{}{appendix}{\secn@m}{\secn@m}}%
\writedef{#1\leftbracket#1}\wrlabeL{#1=#1}}
\newwrite\tfile \def\writetoca#1{}
\def\leaderfill{\leaders\hbox to 1em{\hss.\hss}\hfill}
%	use this to write file with table of contents
\def\writetoc{\immediate\openout\tfile=\jobname.toc
   \def\writetoca##1{{\edef\next{\write\tfile{\noindent ##1
   \string\leaderfill {\string\hyperref{}{page}{\noexpand\number\pageno}%
                       {\noexpand\number\pageno}} \par}}\next}}}
%       and this lists table of contents on second pass
\newread\ch@ckfile
\def\listtoc{\immediate\closeout\tfile\immediate\openin\ch@ckfile=\jobname.toc
\ifeof\ch@ckfile\message{no file \jobname.toc, no table of contents this pass}%
\else\closein\ch@ckfile\centerline{\bf Contents}\nobreak\medskip%
{\baselineskip=12pt\footnotefont\parskip=0pt\catcode`\@=11\input\jobname.toc
\catcode`\@=12\bigbreak\bigskip}\fi}
\catcode`\@=12 % at signs are no longer letters
%
%	Unpleasantness in calling in abstract and title fonts
\edef\tfontsize{\ifx\answ\bigans scaled\magstep3\else scaled\magstep4\fi}
\font\titlerm=cmr10 \tfontsize \font\titlerms=cmr7 \tfontsize
\font\titlermss=cmr5 \tfontsize \font\titlei=cmmi10 \tfontsize
\font\titleis=cmmi7 \tfontsize \font\titleiss=cmmi5 \tfontsize
\font\titlesy=cmsy10 \tfontsize \font\titlesys=cmsy7 \tfontsize
\font\titlesyss=cmsy5 \tfontsize \font\titleit=cmti10 \tfontsize
\skewchar\titlei='177 \skewchar\titleis='177 \skewchar\titleiss='177
\skewchar\titlesy='60 \skewchar\titlesys='60 \skewchar\titlesyss='60
\def\titlefont{\def\rm{\fam0\titlerm}% switch to title font
\textfont0=\titlerm \scriptfont0=\titlerms \scriptscriptfont0=\titlermss
\textfont1=\titlei \scriptfont1=\titleis \scriptscriptfont1=\titleiss
\textfont2=\titlesy \scriptfont2=\titlesys \scriptscriptfont2=\titlesyss
\textfont\itfam=\titleit \def\it{\fam\itfam\titleit}\rm}
 \ifx\answ\bigans\else scaled\magstep1\fi
\ifx\answ\bigans\def\abstractfont{\tenpoint}\else
\font\absit=cmti10 scaled \magstep1
\font\abssl=cmsl10 scaled \magstep1
\font\absrm=cmr10 scaled\magstep1 \font\absrms=cmr7 scaled\magstep1
\font\absrmss=cmr5 scaled\magstep1 \font\absi=cmmi10 scaled\magstep1
\font\absis=cmmi7 scaled\magstep1 \font\absiss=cmmi5 scaled\magstep1
\font\abssy=cmsy10 scaled\magstep1 \font\abssys=cmsy7 scaled\magstep1
\font\abssyss=cmsy5 scaled\magstep1 \font\absbf=cmbx10 scaled\magstep1
\skewchar\absi='177 \skewchar\absis='177 \skewchar\absiss='177
\skewchar\abssy='60 \skewchar\abssys='60 \skewchar\abssyss='60
\def\abstractfont{\def\rm{\fam0\absrm}% switch to abstract font
\textfont0=\absrm \scriptfont0=\absrms \scriptscriptfont0=\absrmss
\textfont1=\absi \scriptfont1=\absis \scriptscriptfont1=\absiss
\textfont2=\abssy \scriptfont2=\abssys \scriptscriptfont2=\abssyss
\textfont\itfam=\absit \def\it{\fam\itfam\absit}\def\footnotefont{\tenpoint}%
\textfont\slfam=\abssl \def\sl{\fam\slfam\abssl}%
\textfont\bffam=\absbf \def\bf
{\fam\bffam\absbf}\rm}\fi
\def\tenpoint{\def\rm{\fam0\tenrm}% switch back to 10-point type
\textfont0=\tenrm \scriptfont0=\sevenrm \scriptscriptfont0=\fiverm
\textfont1=\teni  \scriptfont1=\seveni  \scriptscriptfont1=\fivei
\textfont2=\tensy \scriptfont2=\sevensy \scriptscriptfont2=\fivesy
\textfont\itfam=\tenit \def\it{\fam\itfam\tenit}\def\footnotefont{\ninepoint}%
\textfont\bffam=\tenbf \def\bf{\fam\bffam\tenbf}\def\sl{\fam\slfam\tensl}\rm}
\font\ninerm=cmr9 \font\sixrm=cmr6 \font\ninei=cmmi9 \font\sixi=cmmi6
\font\ninesy=cmsy9 \font\sixsy=cmsy6 \font\ninebf=cmbx9
\font\nineit=cmti9 \font\ninesl=cmsl9 \skewchar\ninei='177
\skewchar\sixi='177 \skewchar\ninesy='60 \skewchar\sixsy='60
\def\ninepoint{\def\rm{\fam0\ninerm}% switch to footnote font
\textfont0=\ninerm \scriptfont0=\sixrm \scriptscriptfont0=\fiverm
\textfont1=\ninei \scriptfont1=\sixi \scriptscriptfont1=\fivei
\textfont2=\ninesy \scriptfont2=\sixsy \scriptscriptfont2=\fivesy
\textfont\itfam=\ninei \def\it{\fam\itfam\nineit}\def\sl{\fam\slfam\ninesl}%
\textfont\bffam=\ninebf \def\bf{\fam\bffam\ninebf}\rm}
%
%---------------------------------------------------------------------
%
\def\noblackbox{\overfullrule=0pt}
\hyphenation{anom-aly anom-alies coun-ter-term coun-ter-terms}
\def\inv{^{\raise.15ex\hbox{${\scriptscriptstyle -}$}\kern-.05em 1}}

\def\Dsl{\,\raise.15ex\hbox{/}\mkern-13.5mu D} %this one can be subscripted
\def\dsl{\raise.15ex\hbox{/}\kern-.57em\partial}

\def\tr{{\rm tr}} \def\Tr{{\rm Tr}}
 %pound sterling
\def\lspace{\ifx\answ\bigans{}\else\qquad\fi}
\def\lbspace{\ifx\answ\bigans{}\else\hskip-.2in\fi} % $$\lbspace...$$

\def\boxeqn#1{\vcenter{\vbox{\hrule\hbox{\vrule\kern3pt\vbox{\kern3pt
	\hbox{${\displaystyle #1}$}\kern3pt}\kern3pt\vrule}\hrule}}}
\def\mbox#1#2{\vcenter{\hrule \hbox{\vrule height#2in
		\kern#1in \vrule} \hrule}}  %e.g. \mbox{.1}{.1}
%	matters of taste
%\def\tilde{\widetilde} \def\bar{\overline} \def\hat{\widehat}
%
% some sample definitions
  %     curly letters
\def\CA{{\cal A}} \def\CC{{\cal C}}  
\def\CL{{\cal L}}  \def\CI{{\cal I}} 
   \def\CT{{\cal T}}

\def\darr#1{\raise1.5ex\hbox{$\leftrightarrow$}\mkern-16.5mu #1}
 %pound sterling

\def\half{{\textstyle{1\over2}}} %puts a small half in a displayed eqn
\def\roughly#1{\raise.3ex\hbox{$#1$\kern-.75em\lower1ex\hbox{$\sim$}}}

\def\bb{
\font\tenmsb=msbm10
\font\sevenmsb=msbm7
\font\fivemsb=msbm5
\textfont1=\tenmsb
\scriptfont1=\sevenmsb
\scriptscriptfont1=\fivemsb
}

\input amssym

\input epsf

\def\IZ{\relax\ifmmode\mathchoice
{\hbox{\cmss Z\kern-.4em Z}}{\hbox{\cmss Z\kern-.4em Z}} {\lower.9pt\hbox{\cmsss Z\kern-.4em Z}}
{\lower1.2pt\hbox{\cmsss Z\kern-.4em Z}}\else{\cmss Z\kern-.4em Z}\fi}

\newif\ifdraft\draftfalse
%\drafttrue
\newif\ifinter\interfalse
%\intertrue
\ifdraft\draftmode\else\interfalse\fi
\def\journal#1&#2(#3){\unskip, \sl #1\ \bf #2 \rm(19#3) }
\def\andjournal#1&#2(#3){\sl #1~\bf #2 \rm (19#3) }

\def\frac#1#2{{#1\over#2}}

\def\half{\frac12}

\def\inbar{\,\vrule height1.5ex width.4pt depth0pt}
\def\IC{\relax\hbox{$\inbar\kern-.3em{\rm C}$}}
\def\IR{\relax{\rm I\kern-.18em R}}
\def\IP{\relax{\rm I\kern-.18em P}}
%\def\Z{{\bf Z}}

%
%%%%%%%%%%%%%%%%%%%%%%%%%%%%%%%%%%%%
%

%\def\ap#1#2#3{Ann. Phys. {\bf #1} (#2) #3}

%
\catcode`\@=11
\def\slash#1{\mathord{\mathpalette\c@ncel{#1}}}
\overfullrule=0pt

\def\CC{{\cal C}}

\def\Z{\hbox{$\bb Z$}}

\def\underrel#1\over#2{\mathrel{\mathop{\kern\z@#1}\limits_{#2}}}

\catcode`\@=12

%%%%%%%%%%%%%%%%%%%%%%%%%%%%%%%%%%%%%%%%%%%%%%%%%%%%%%%%%%%%%%

%

\def\det{{\rm det}}
\def\tr{{\rm tr}}
\def\mod{{\rm mod}}

\def\det{{\rm det}}
\def\exp{{\rm exp}}

%%%%%%%%%%%%%%%%%%%%%%%%%%%%%%%%%%%%%%%%%%%%%%%%%%%%%%%%%%%%%%
% new defs:

\def\[{[}
\def\]{]}

\def\comment#1{ }

%%%%%%%%%%%%%%%%%%%%%%%%%%%%%%%%%%%%%%%%%%%%%%%%%%%%%%%%%%%%%%
%%% Oskar's definitions:
%
%% A box for a short draft note
\def\draftnote#1{\ifdraft{\baselineskip2ex
                 \vbox{\kern1em\hrule\hbox{\vrule\kern1em\vbox{\kern1ex
                 \noindent \underbar{NOTE}: #1
             \vskip1ex}\kern1em\vrule}\hrule}}\fi}
%% A box for a short internal note
\def\internote#1{\ifinter{\baselineskip2ex
                 \vbox{\kern1em\hrule\hbox{\vrule\kern1em\vbox{\kern1ex
                 \noindent \underbar{Internal Note}: #1
             \vskip1ex}\kern1em\vrule}\hrule}}\fi}
%% A few internal words

%
%% Greek letters
%
%\def\al{\alpha}
%\def\bt{\beta}
%\def\gm{\gamma}                \def\Gm{\Gamma}
%\def\dl{\delta}                \def\Dl{\Delta}
%\def\ep{\epsilon}
%\def\vep{\varepsilon}

%\def\io{\iota}
%\def\kp{\kappa}
%\def\lm{\lambda}               \def\Lm{\Lambda}
%%\mu,\nu unchanged

%\def\th{\theta}               \def\Th{\Theta}
%\def\vth{\vartheta}
%%\phi unchanged               \Phi unchanged
%\def\vph{\varphi}
%%\psi unchanged               \Psi unchanged
%%\chi unchanged

%\def\om{\omega}               \def\Om{\Omega}
%%\pi unchanged                \Pi unchanged
%\def\vpi{\varpi}
%%\rho unchanged
%\def\vro{\varrho}
%\def\sg{\sigma}               \def\Sg{\Sigma}
%\def\vsg{\varsigma}
%%\tau unchanged
%\def\up{\upsilon}             \Up{\Upsilon}
%%\xi unchanged                \Xi unchanged
%%\eta unchanged
%\def\zt{\zeta}
%
%% Capital roman double letters (blackboard font)
%
%\def\inbar{\hskip.3em\vrule height1.5ex width.4pt depth0pt}
%\def\IC{\relax{\inbar\kern-.3em{\rm C}}}
%\def\IN{\relax{\rm I\kern-.16em N}}
%\def\IP{\relax{\rm I\kern-.18em P}}
%\def\IQ{\relax\hbox{$\inbar$\kern-.3em{\rm Q}}}
%\def\IR{\relax{\rm I\kern-.18em R}}
%\def\IZ{\relax{\rm Z\kern-.8em Z}}
%
%% Other Defs
%

%

\def\inv{^{-1}}

%mydef

\def\tr{{\rm tr}}
\def\Tr{{\rm Tr}}

\def\b{\beta}

%\SonXQA
\lref\SonXQA{
  D.~T.~Son,
  ``Is the Composite Fermion a Dirac Particle?,''
Phys.\ Rev.\ X {\bf 5}, no. 3, 031027 (2015).
[arXiv:1502.03446 [cond-mat.mes-hall]].
%%CITATION = EFI-15-6%%
}

%\FradkinTT
\lref\FradkinTT{
  E.~H.~Fradkin and F.~A.~Schaposnik,
  ``The Fermion - boson mapping in three-dimensional quantum field theory,''
Phys.\ Lett.\ B {\bf 338}, 253 (1994).
[hep-th/9407182].
%%CITATION = hep-th/9407182%%
}

%\PolyakovMD
\lref\PolyakovMD{
  A.~M.~Polyakov,
  ``Fermi-Bose Transmutations Induced by Gauge Fields,''
Mod.\ Phys.\ Lett.\ A {\bf 3}, 325 (1988).
%%CITATION = Print-88-0037 (LANDAU)%%
}

%\WittenXI
\lref\WittenXI{
  E.~Witten,
  ``The Verlinde algebra and the cohomology of the Grassmannian,''
In *Cambridge 1993, Geometry, topology, and physics* 357-422.
[hep-th/9312104].
%%CITATION = hep-th/9312104%%
}

%\ShajiIS
\lref\ShajiIS{
  N.~Shaji, R.~Shankar and M.~Sivakumar,
  ``On Bose-fermi Equivalence in a U(1) Gauge Theory With {Chern-Simons} Action,''
Mod.\ Phys.\ Lett.\ A {\bf 5}, 593 (1990).
%%CITATION = IMSc/89/18%%
}

%\WangQMT
\lref\WangQMT{
  C.~Wang and T.~Senthil,
  ``Dual Dirac Liquid on the Surface of the Electron Topological Insulator,''
Phys.\ Rev.\ X {\bf 5}, no. 4, 041031 (2015). [arXiv:1505.05141 [cond-mat.str-el]].
}

%\MetlitskiEKA
\lref\MetlitskiEKA{
  M.~A.~Metlitski and A.~Vishwanath,
  ``Particle-vortex duality of 2d Dirac fermion from electric-magnetic duality of 3d topological insulators,''
[arXiv:1505.05142 [cond-mat.str-el]].
%%CITATION = arXiv:1505.05142%%
}

%\FidkowskiJUA
\lref\FidkowskiJUA{
  L.~Fidkowski, X.~Chen and A.~Vishwanath,
  ``Non-Abelian Topological Order on the Surface of a 3D Topological Superconductor from an Exactly Solved Model,''
Phys.\ Rev.\ X {\bf 3}, no. 4, 041016 (2013).
[arXiv:1305.5851 [cond-mat.str-el]].
%%CITATION = arXiv:1305.5851%%
}

%\WangFQL
\lref\WangFQL{
  C.~Wang and T.~Senthil,
  ``Half-filled Landau level, topological insulator surfaces, and three-dimensional quantum spin liquids,''
Phys.\ Rev.\ B {\bf 93}, no. 8, 085110 (2016). [arXiv:1507.08290 [cond-mat.st-el]].
}

%\GeraedtsPVA
\lref\GeraedtsPVA{
  S.~D.~Geraedts, M.~P.~Zaletel, R.~S.~K.~Mong, M.~A.~Metlitski, A.~Vishwanath and O.~I.~Motrunich,
  ``The half-filled Landau level: the case for Dirac composite fermions,''
Science {\bf 352}, 197 (2016).
[arXiv:1508.04140 [cond-mat.str-el]].
%%CITATION = arXiv:1508.04140%%
}

%\MrossIDY
\lref\MrossIDY{
  D.~F.~Mross, J.~Alicea and O.~I.~Motrunich,
  ``Explicit derivation of duality between a free Dirac cone and quantum electrodynamics in (2+1) dimensions,''
[arXiv:1510.08455 [cond-mat.str-el]].
%%CITATION = arXiv:1510.08455%%
}

%\RoscherWOX
\lref\RoscherWOX{
  D.~Roscher, E.~Torres and P.~Strack,
  ``Dual QED$_3$ at "$N_F = 1/2$" is an interacting CFT in the infrared,''
[arXiv:1605.05347 [cond-mat.str-el]].
%%CITATION = arXiv:1605.05347%%
}

%\WangCTO
\lref\WangCTO{
  C.~Wang and T.~Senthil,
  ``Time-Reversal Symmetric $U(1)$ Quantum Spin Liquids,''
Phys.\ Rev.\ X {\bf 6}, no. 1, 011034 (2016).
}

%\MetlitskiYQA
\lref\MetlitskiYQA{
  M.~A.~Metlitski,
  ``$S$-duality of $u(1)$ gauge theory with $\theta =\pi$ on non-orientable manifolds: Applications to topological insulators and superconductors,''
[arXiv:1510.05663 [hep-th]].
%%CITATION = arXiv:1510.05663%%
}

%\MulliganGLM
\lref\MulliganGLM{
  M.~Mulligan, S.~Raghu and M.~P.~A.~Fisher,
  ``Emergent particle-hole symmetry in the half-filled Landau level,''
[arXiv:1603.05656 [cond-mat.str-el]].
%%CITATION = arXiv:1603.05656%%
}

%\WangGQJ
\lref\WangGQJ{
  C.~Wang and T.~Senthil,
  ``Composite fermi liquids in the lowest Landau level,''
[arXiv:1604.06807 [cond-mat.str-el]].
%%CITATION = arXiv:1604.06807%%
}

%\WilczekDU
\lref\WilczekDU{
  F.~Wilczek,
  ``Magnetic Flux, Angular Momentum, and Statistics,''
Phys.\ Rev.\ Lett.\  {\bf 48}, 1144 (1982).
%%CITATION = NSF-ITP-81-117%%
}
%\JainTX
\lref\JainTX{
  J.~K.~Jain,
  ``Composite fermion approach for the fractional quantum Hall effect,''
Phys.\ Rev.\ Lett.\  {\bf 63}, 199 (1989).
}

%\BorokhovCG
\lref\BorokhovCG{
  V.~Borokhov, A.~Kapustin and X.~k.~Wu,
  ``Monopole operators and mirror symmetry in three-dimensions,''
JHEP {\bf 0212}, 044 (2002).
[hep-th/0207074].
%%CITATION = hep-th/0207074%%
}

%\AharonyBX
\lref\AharonyBX{
  O.~Aharony, A.~Hanany, K.~A.~Intriligator, N.~Seiberg and M.~J.~Strassler,
  ``Aspects of N=2 supersymmetric gauge theories in three-dimensions,''
Nucl.\ Phys.\ B {\bf 499}, 67 (1997).
[hep-th/9703110].
%%CITATION = hep-th/9703110%%
}

%\PeskinKP
\lref\PeskinKP{
  M.~E.~Peskin,
  ``Mandelstam 't Hooft Duality in Abelian Lattice Models,''
Annals Phys.\  {\bf 113}, 122 (1978).
%%CITATION = HUTP-77/A083%%
}
%\DasguptaZZ
\lref\DasguptaZZ{
  C.~Dasgupta and B.~I.~Halperin,
  ``Phase Transition in a Lattice Model of Superconductivity,''
Phys.\ Rev.\ Lett.\  {\bf 47}, 1556 (1981).
}

%\JainGZA
\lref\JainGZA{
  S.~Jain, S.~Minwalla and S.~Yokoyama,
  ``Chern Simons duality with a fundamental boson and fermion,''
JHEP {\bf 1311}, 037 (2013).
[arXiv:1305.7235 [hep-th]].
%%CITATION = TIFR-TH-13-17%%
}

%\KachruRMA
\lref\KachruRMA{
  S.~Kachru, M.~Mulligan, G.~Torroba and H.~Wang,
  ``Mirror symmetry and the half-filled Landau level,''
Phys.\ Rev.\ B {\bf 92}, 235105 (2015).
[arXiv:1506.01376 [cond-mat.str-el]].
%%CITATION = arXiv:1506.01376%%
}

%\ClossetVP
\lref\ClossetVP{
  C.~Closset, T.~T.~Dumitrescu, G.~Festuccia, Z.~Komargodski and N.~Seiberg,
  ``Comments on Chern-Simons Contact Terms in Three Dimensions,''
JHEP {\bf 1209}, 091 (2012).
[arXiv:1206.5218 [hep-th]].
%%CITATION = PUTP-2417%%
}

%\Gur-AriPCA
\lref\GurPCA{
  G.~Gur-Ari and R.~Yacoby,
  ``Three Dimensional Bosonization From Supersymmetry,''
JHEP {\bf 1511}, 013 (2015).
[arXiv:1507.04378 [hep-th]].
%%CITATION = PUPT-2482%%
}
%\RadicevicYLA
\lref\RadicevicYLA{
  D.~Radicevic,
  ``Disorder Operators in Chern-Simons-Fermion Theories,''
JHEP {\bf 1603}, 131 (2016).
[arXiv:1511.01902 [hep-th]].
%%CITATION = SU-ITP-15-16%%
}

%\NaculichPA
\lref\NaculichPA{
  S.~G.~Naculich, H.~A.~Riggs and H.~J.~Schnitzer,
  ``Group Level Duality in {WZW} Models and {Chern-Simons} Theory,''
Phys.\ Lett.\ B {\bf 246}, 417 (1990).
%%CITATION = BRX-TH-293%%
}

%\MlawerUV
\lref\MlawerUV{
  E.~J.~Mlawer, S.~G.~Naculich, H.~A.~Riggs and H.~J.~Schnitzer,
  ``Group level duality of WZW fusion coefficients and Chern-Simons link observables,''
Nucl.\ Phys.\ B {\bf 352}, 863 (1991).
%%CITATION = BRX-TH-301%%
}

%\BorokhovIB
\lref\BorokhovIB{
  V.~Borokhov, A.~Kapustin and X.~k.~Wu,
  ``Topological disorder operators in three-dimensional conformal field theory,''
JHEP {\bf 0211}, 049 (2002).
[hep-th/0206054].
%%CITATION = hep-th/0206054%%
}

%\NakanishiHJ
\lref\NakanishiHJ{
  T.~Nakanishi and A.~Tsuchiya,
  ``Level rank duality of WZW models in conformal field theory,''
Commun.\ Math.\ Phys.\  {\bf 144}, 351 (1992).
%%CITATION = NU-MATH-002%%
}

%\GiveonZN
\lref\GiveonZN{
  A.~Giveon and D.~Kutasov,
  ``Seiberg Duality in Chern-Simons Theory,''
Nucl.\ Phys.\ B {\bf 812}, 1 (2009).
[arXiv:0808.0360 [hep-th]].
%%CITATION = arXiv:0808.0360%%
}
%\BeniniMF
\lref\BeniniMF{
  F.~Benini, C.~Closset and S.~Cremonesi,
  ``Comments on 3d Seiberg-like dualities,''
JHEP {\bf 1110}, 075 (2011).
[arXiv:1108.5373 [hep-th]].
%%CITATION = arXiv:1108.5373%%
}
%\AharonyDHA
\lref\AharonyDHA{
  O.~Aharony, S.~S.~Razamat, N.~Seiberg and B.~Willett,
  ``3d dualities from 4d dualities,''
JHEP {\bf 1307}, 149 (2013).
[arXiv:1305.3924 [hep-th]].
%%CITATION = WIS-04-13-APR-DPPA%%
}

%\AharonyJZ
\lref\AharonyJZ{
  O.~Aharony, G.~Gur-Ari and R.~Yacoby,
  ``d=3 Bosonic Vector Models Coupled to Chern-Simons Gauge Theories,''
JHEP {\bf 1203}, 037 (2012).
[arXiv:1110.4382 [hep-th]].
%%CITATION = arXiv:1110.4382%%
}
%\GiombiKC
\lref\GiombiKC{
  S.~Giombi, S.~Minwalla, S.~Prakash, S.~P.~Trivedi, S.~R.~Wadia and X.~Yin,
  ``Chern-Simons Theory with Vector Fermion Matter,''
Eur.\ Phys.\ J.\ C {\bf 72}, 2112 (2012).
[arXiv:1110.4386 [hep-th]].
%%CITATION = arXiv:1110.4386%%
}
%\AharonyNH
\lref\AharonyNH{
  O.~Aharony, G.~Gur-Ari and R.~Yacoby,
  ``Correlation Functions of Large N Chern-Simons-Matter Theories and Bosonization in Three Dimensions,''
JHEP {\bf 1212}, 028 (2012).
[arXiv:1207.4593 [hep-th]].
%%CITATION = WIS-13-12-JUL-DPPA%%
}

%\VasilievVF
\lref\VasilievVF{
  M.~A.~Vasiliev,
  ``Holography, Unfolding and Higher-Spin Theory,''
J.\ Phys.\ A {\bf 46}, 214013 (2013).
[arXiv:1203.5554 [hep-th]].
%%CITATION = FIAN-TD-06-12%%
}

%\FradkinTT
\lref\FradkinTT{
  E.~H.~Fradkin and F.~A.~Schaposnik,
  ``The Fermion - boson mapping in three-dimensional quantum field theory,''
Phys.\ Lett.\ B {\bf 338}, 253 (1994).
[hep-th/9407182].
%%CITATION = hep-th/9407182%%
}

%\PolyakovMD
\lref\PolyakovMD{
  A.~M.~Polyakov,
  ``Fermi-Bose Transmutations Induced by Gauge Fields,''
Mod.\ Phys.\ Lett.\ A {\bf 3}, 325 (1988).
%%CITATION = Print-88-0037 (LANDAU)%%
}

%\ShajiIS
\lref\ShajiIS{
  N.~Shaji, R.~Shankar and M.~Sivakumar,
  ``On Bose-fermi Equivalence in a U(1) Gauge Theory With {Chern-Simons} Action,''
Mod.\ Phys.\ Lett.\ A {\bf 5}, 593 (1990).
%%CITATION = IMSc/89/18%%
}

%\AlvarezGaumeNF
\lref\AlvarezGaumeNF{
  L.~Alvarez-Gaume, S.~Della Pietra and G.~W.~Moore,
  ``Anomalies and Odd Dimensions,''
Annals Phys.\  {\bf 163}, 288 (1985).
%%CITATION = HUTP-84-A028%%
}

%\AharonyMJS
\lref\AharonyMJS{
  O.~Aharony,
  ``Baryons, monopoles and dualities in Chern-Simons-matter theories,''
JHEP {\bf 1602}, 093 (2016).
[arXiv:1512.00161 [hep-th]].
%%CITATION = WIS-12-15-NOV-DPPA%%
}

%\DaiKQ
\lref\DaiKQ{
  X.~z.~Dai and D.~S.~Freed,
  ``eta invariants and determinant lines,''
J.\ Math.\ Phys.\  {\bf 35}, 5155 (1994), Erratum: [J.\ Math.\ Phys.\  {\bf 42}, 2343 (2001)].
[hep-th/9405012].
%%CITATION = hep-th/9405012%%
}

%\WittenABA
\lref\WittenABA{
  E.~Witten,
  ``Fermion Path Integrals And Topological Phases,''
[arXiv:1508.04715 [cond-mat.mes-hall]].
%%CITATION = arXiv:1508.04715%%
}

%\SeibergRSG
\lref\SeibergRSG{
  N.~Seiberg and E.~Witten,
  ``Gapped Boundary Phases of Topological Insulators via Weak Coupling,''
[arXiv:1602.04251 [cond-mat.str-el]].
%%CITATION = arXiv:1602.04251%%
}

%\WittenYA
\lref\WittenYA{
  E.~Witten,
  ``SL(2,Z) action on three-dimensional conformal field theories with Abelian symmetry,''
In *Shifman, M. (ed.) et al.: From fields to strings, vol. 2* 1173-1200.
[hep-th/0307041].
%%CITATION = hep-th/0307041%%
}

%\HalperinMH
\lref\HalperinMH{
  B.~I.~Halperin, P.~A.~Lee and N.~Read,
  ``Theory of the half filled Landau level,''
Phys.\ Rev.\ B {\bf 47}, 7312 (1993).
%%CITATION = PRINT-92-0424%%
}
%\DouglasEX
\lref\DouglasEX{
  M.~R.~Douglas,
  ``Chern-Simons-Witten theory as a topological Fermi liquid,''
[hep-th/9403119].
%%CITATION = hep-th/9403119%%
}
%\PotterCDN
\lref\PotterCDN{
  A.~C.~Potter, M.~Serbyn and A.~Vishwanath,
  ``Thermoelectric transport signatures of Dirac composite fermions in the half-filled Landau level,''
[arXiv:1512.06852 [cond-mat.str-el]].
%%CITATION = arXiv:1512.06852%%
}

%\WangUKY
\lref\WangUKY{
  C.~Wang, A.~C.~Potter and T.~Senthil,
  ``Gapped symmetry preserving surface state for the electron topological insulator,''
Phys.\ Rev.\ B {\bf 88}, no. 11, 115137 (2013).
[1306.3223].
%%CITATION = 1306.3223%%
}

%\MetlitskiBPA
\lref\MetlitskiBPA{
  M.~A.~Metlitski, C.~L.~Kane and M.~P.~A.~Fisher,
  ``Symmetry-respecting topologically ordered surface phase of three-dimensional electron topological insulators,''
Phys.\ Rev.\ B {\bf 92}, no. 12, 125111 (2015).
}

%\ChenJHA
\lref\ChenJHA{
  X.~Chen, L.~Fidkowski and A.~Vishwanath,
  ``Symmetry Enforced Non-Abelian Topological Order at the Surface of a Topological Insulator,''
Phys.\ Rev.\ B {\bf 89}, no. 16, 165132 (2014).
[arXiv:1306.3250 [cond-mat.str-el]].
%%CITATION = arXiv:1306.3250%%
}

%\KarchSXI
\lref\KarchSXI{
  A.~Karch and D.~Tong,
  ``Particle-Vortex Duality from 3d Bosonization,''
[arXiv: 1606.01893 [hep-th]].
%%CITATION = arXiv:1606.01893%%
}
%\MuruganZAL
\lref\MuruganZAL{
  J.~Murugan and H.~Nastase,
  ``Particle-vortex duality in topological insulators and superconductors,''
[arXiv:1606.01912 [hep-th]].
%%CITATION = arXiv:1606.01912%%
}

%\BondersonPLA
\lref\BondersonPLA{
  P.~Bonderson, C.~Nayak and X.~L.~Qi,
  ``A time-reversal invariant topological phase at the surface of a 3D topological insulator,''
J.\ Stat.\ Mech.\  {\bf 2013}, P09016 (2013).
}

%\SenthilJK
\lref\SenthilJK{
  T.~Senthil and M.~P.~A.~Fisher,
  ``Competing orders, non-linear sigma models, and topological terms in quantum magnets,''
Phys.\ Rev.\ B {\bf 74}, 064405 (2006).
[cond-mat/0510459].
%%CITATION = cond-mat/0510459%%
}

%\WittenGF
\lref\WittenGF{
  E.~Witten,
  ``On S duality in Abelian gauge theory,''
Selecta Math.\  {\bf 1}, 383 (1995).
[hep-th/9505186].
%%CITATION = hep-th/9505186%%
}

%\MetlitskiYQA
\lref\MetlitskiYQA{
  M.~A.~Metlitski,
  ``$S$-duality of $u(1)$ gauge theory with $\theta =\pi$ on non-orientable manifolds: Applications to topological insulators and superconductors,''
[arXiv:1510.05663 [hep-th]].
%%CITATION = arXiv:1510.05663%%
}

%\IntriligatorEX
\lref\IntriligatorEX{
  K.~A.~Intriligator and N.~Seiberg,
  ``Mirror symmetry in three-dimensional gauge theories,''
Phys.\ Lett.\ B {\bf 387}, 513 (1996).
[hep-th/9607207].
%%CITATION = hep-th/9607207%%
}

%\KapustinHA
\lref\KapustinHA{
  A.~Kapustin and M.~J.~Strassler,
  ``On mirror symmetry in three-dimensional Abelian gauge theories,''
JHEP {\bf 9904}, 021 (1999).
[hep-th/9902033].
%%CITATION = hep-th/9902033%%
}

%\GaiottoAK
\lref\GaiottoAK{
  D.~Gaiotto and E.~Witten,
  ``S-Duality of Boundary Conditions In N=4 Super Yang-Mills Theory,''
Adv.\ Theor.\ Math.\ Phys.\  {\bf 13}, no. 3, 721 (2009).
[arXiv:0807.3720 [hep-th]].
%%CITATION = arXiv:0807.3720%%
}

%\WittenYA
\lref\WittenYA{
  E.~Witten,
  ``SL(2,Z) action on three-dimensional conformal field theories with Abelian symmetry,''
In *Shifman, M. (ed.) et al.: From fields to strings, vol. 2* 1173-1200.
[hep-th/0307041].
%%CITATION = hep-th/0307041%%
}
%\SeibergPQ
\lref\SeibergPQ{
  N.~Seiberg,
  ``Electric - magnetic duality in supersymmetric nonAbelian gauge theories,''
Nucl.\ Phys.\ B {\bf 435}, 129 (1995).
[hep-th/9411149].
%%CITATION = hep-th/9411149%%
}

\lref\MPS{C. Wang, A. C. Potter, and T. Senthil,
``Classification Of Interacting Electronic Topological Insulators In Three Dimensions,'' Science {\bf 343} (2014) 629, arXiv:1306.3238.}

\lref\McG{S. M. Kravec, J. McGreevy, and B. Swingle, ``All-Fermion Electrodynamics And Fermion Number Anomaly Inflow,'' arXiv:1409.8339.}

%\BarkeshliIDA
\lref\BarkeshliIDA{
  M.~Barkeshli and J.~McGreevy,
  ``Continuous transition between fractional quantum Hall and superfluid states,''
Phys.\ Rev.\ B {\bf 89}, no. 23, 235116 (2014).
}

%\ChenCD
\lref\ChenCD{
  W.~Chen, M.~P.~A.~Fisher and Y.~S.~Wu,
  ``Mott transition in an anyon gas,''
Phys.\ Rev.\ B {\bf 48}, 13749 (1993).
[cond-mat/9301037].
%%CITATION = cond-mat/9301037%%
}

%\WangLCA
\lref\WangLCA{
  C.~Wang and T.~Senthil,
  ``Interacting fermionic topological insulators/superconductors in three dimensions,''
Phys.\ Rev.\ B {\bf 89}, no. 19, 195124 (2014), Erratum: [Phys.\ Rev.\ B {\bf 91}, no. 23, 239902 (2015)].
[arXiv:1401.1142 [cond-mat.str-el]].
%%CITATION = arXiv:1401.1142%%
}

%\NakaharaNW
\lref\NakaharaNW{
  M.~Nakahara,
  ``Geometry, topology and physics,''
Boca Raton, USA: Taylor and Francis (2003) 573 p.
}

%\NguyenZN
\lref\NguyenZN{
  A.~K.~Nguyen and A.~Sudbo,
  ``Topological phase fluctuations, amplitude fluctuations, and criticality in extreme type II superconductors,''
Phys.\ Rev.\ B {\bf 60}, 15307 (1999).
[cond-mat/9907385].
%%CITATION = cond-mat/9907385%%
}

%\KajantieVY
\lref\KajantieVY{
  K.~Kajantie, M.~Laine, T.~Neuhaus, A.~Rajantie and K.~Rummukainen,
  ``Duality and scaling in three-dimensional scalar electrodynamics,''
Nucl.\ Phys.\ B {\bf 699}, 632 (2004).
[hep-lat/0402021].
%%CITATION = hep-lat/0402021%%
}

%\GoddardVK
\lref\GoddardVK{
  P.~Goddard, A.~Kent and D.~I.~Olive,
  ``Virasoro Algebras and Coset Space Models,''
Phys.\ Lett.\ B {\bf 152}, 88 (1985).
%%CITATION = DAMTP-84-22%%
}

%\ClossetVP
\lref\ClossetVP{
  C.~Closset, T.~T.~Dumitrescu, G.~Festuccia, Z.~Komargodski and N.~Seiberg,
  ``Comments on Chern-Simons Contact Terms in Three Dimensions,''
JHEP {\bf 1209}, 091 (2012).
[arXiv:1206.5218 [hep-th]].
%%CITATION = PUTP-2417%%
}

%\GaiottoKFA
\lref\GaiottoKFA{
  D.~Gaiotto, A.~Kapustin, N.~Seiberg and B.~Willett,
  ``Generalized Global Symmetries,''
JHEP {\bf 1502}, 172 (2015).
[arXiv:1412.5148 [hep-th]].
%%CITATION = arXiv:1412.5148%%
}

%\AlvarezGaumeNF
\lref\AlvarezGaumeNF{
  L.~Alvarez-Gaume, S.~Della Pietra and G.~W.~Moore,
  ``Anomalies and Odd Dimensions,''
Annals Phys.\  {\bf 163}, 288 (1985).
%%CITATION = HUTP-84-A028%%
}
%\WittenABA
\lref\WittenABA{
  E.~Witten,
  ``Fermion Path Integrals And Topological Phases,''
[arXiv:1508.04715 [cond-mat.mes-hall]].
%%CITATION = arXiv:1508.04715%%
}

\def\CSg{{\rm CS_{grav}}}

% black board bold

\def\bb{
\font\tenmsb=msbm10
\font\sevenmsb=msbm7
\font\fivemsb=msbm5
\textfont1=\tenmsb
\scriptfont1=\sevenmsb
\scriptscriptfont1=\fivemsb
}

% MACROS

% bras, kets, ...

% tilde, hat, bar, ...

\def\tilde{\widetilde}

\def\hat{\widehat}

\def\bar{\overline}
\def\b{\bar}
\def\bsq#1{{{\b{#1}}^{\lower 2.5pt\hbox{$\scriptstyle 2$}}}}
\def\bexp#1#2{{{\b{#1}}^{\lower 2.5pt\hbox{$\scriptstyle #2$}}}}
\def\dotexp#1#2{{{#1}^{\lower 2.5pt\hbox{$\scriptstyle #2$}}}}

% basic math

\def\rt2{\sqrt{2}}
\def\half {{1 \over 2}}

\def\mod{{\rm mod}}
\def\det{\mathop{\rm det}}

\def\Tr{\mathop{\rm Tr}}
\def\tr{\mathop{\rm tr}}

% dotted spinor indices

% capital cal letters

\def\CA{{\cal A}}

\def\CC{{\cal C}}

\def\CI{{\cal I}}

\def\CL{{\cal L}}

\def\CN{{\cal N}}

\def\CT{{\cal T}}

% double stroke symbols: unit matrix, reals, complex, quaternions, integers, natural numbers

\def\1{{\ds 1}}

\def\Z{\hbox{$\bb Z$}}

% miscellaneous objects

\noblackbox

\def\unit{\relax{\rm 1\kern-.26em I}}
\def\nada{\relax{\rm 0\kern-.30em l}}
\def\tilde{\widetilde}

\def\mod{{\rm mod}}
%\def\det{{\rm det}}

%% MACROS
\noblackbox
\def\IL{\relax{\rm I\kern-.18em L}}
\def\IH{\relax{\rm I\kern-.18em H}}
\def\IR{\relax{\rm I\kern-.18em R}}
\def\IC{\relax\hbox{$\inbar\kern-.3em{\rm C}$}}
\def\IZ{\relax\ifmmode\mathchoice
{\hbox{\cmss Z\kern-.4em Z}}{\hbox{\cmss Z\kern-.4em Z}} {\lower.9pt\hbox{\cmsss Z\kern-.4em Z}}
{\lower1.2pt\hbox{\cmsss Z\kern-.4em Z}}\else{\cmss Z\kern-.4em Z}\fi}

\def\CN {{\cal N}}

\def\partialslash{\not{\hbox{\kern-2pt $\partial$}}}

\def\CL {{\cal L}}

\def\CC {{\cal C}}

\def\CA{{\cal A}}

%% MORE MACROS

\def\CN {{\cal N}}

\def\Tr{{\rm Tr\, }}

\font\manual=manfnt \def\dbend{\lower3.5pt\hbox{\manual\char127}}

\def\IZ{\relax\ifmmode\mathchoice
{\hbox{\cmss Z\kern-.4em Z}}{\hbox{\cmss Z\kern-.4em Z}} {\lower.9pt\hbox{\cmsss Z\kern-.4em Z}}
{\lower1.2pt\hbox{\cmsss Z\kern-.4em Z}}\else{\cmss Z\kern-.4em Z}\fi}
\def\half {{1\over 2}}

\def\bar{\overline}

\def\rt2{\sqrt{2}}
\def\irt2{{1\over\sqrt{2}}}

\def\hat{\widehat}
%  \slashchar puts a slash through a character to represent contraction
%  with Dirac matrices. Use \not instead for negation of relations, and use
%  \hbar for hbar.
\def\slashchar#1{\setbox0=\hbox{$#1$}           % set a box for #1
   \dimen0=\wd0                                 % and get its size
   \setbox1=\hbox{/} \dimen1=\wd1               % get size of /
   \ifdim\dimen0>\dimen1                        % #1 is bigger
      \rlap{\hbox to \dimen0{\hfil/\hfil}}      % so center / in box
      #1                                        % and print #1
   \else                                        % / is bigger
      \rlap{\hbox to \dimen1{\hfil$#1$\hfil}}   % so center #1
      /                                         % and print /
   \fi}

%%% Yuji's macros

\def\figcaption#1#2{\DefWarn#1\xdef#1{Figure~\noexpand\hyperref{}{figure}%
{\the\figno}{\the\figno}}\writedef{#1\leftbracket Figure\noexpand~\xfig#1}%
\medskip\centerline{{\footnotefont\bf Figure~\hyperdef\hypernoname{figure}{\the\figno}{\the\figno}:}  #2 \wrlabeL{#1=#1}}%
\global\advance\figno by1}

%\MaldacenaSS
\lref\MaldacenaSS{
  J.~M.~Maldacena, G.~W.~Moore and N.~Seiberg,
  ``D-brane charges in five-brane backgrounds,''
JHEP {\bf 0110}, 005 (2001).
[hep-th/0108152].
%%CITATION = hep-th/0108152%%
}

%\BanksZN
\lref\BanksZN{
  T.~Banks and N.~Seiberg,
  ``Symmetries and Strings in Field Theory and Gravity,''
Phys.\ Rev.\ D {\bf 83}, 084019 (2011).
[arXiv:1011.5120 [hep-th]].
%%CITATION = arXiv:1011.5120%%
}

%\SeibergRSG
\lref\SeibergRSG{
  N.~Seiberg and E.~Witten,
  ``Gapped Boundary Phases of Topological Insulators via Weak Coupling,''
[arXiv:1602.04251 [cond-mat.str-el]].
%%CITATION = arXiv:1602.04251%%
}

%\KapustinGUA
\lref\KapustinGUA{
  A.~Kapustin and N.~Seiberg,
  ``Coupling a QFT to a TQFT and Duality,''
JHEP {\bf 1404}, 001 (2014).
[arXiv:1401.0740 [hep-th]].
%%CITATION = arXiv:1401.0740%%
}

%\MooreYH
\lref\MooreYH{
  G.~W.~Moore and N.~Seiberg,
  ``Taming the Conformal Zoo,''
Phys.\ Lett.\ B {\bf 220}, 422 (1989).
%%CITATION = IASSNS-HEP-89/6%%
}

%\SeibergGMD
\lref\SeibergGMD{
  N.~Seiberg, T.~Senthil, C.~Wang and E.~Witten,
  ``A Duality Web in 2+1 Dimensions and Condensed Matter Physics,''
[arXiv:1606.01989 [hep-th]].
%%CITATION = arXiv:1606.01989%%
}

%\XuNXA
\lref\XuNXA{
  F.~Xu,
  ``Algebraic coset conformal field theories,''
Commun.\ Math.\ Phys.\  {\bf 211}, 1 (2000).
[math/9810035].
%%CITATION = math/9810035%%
}

%\AharonyMJS
\lref\AharonyMJS{
  O.~Aharony,
  ``Baryons, monopoles and dualities in Chern-Simons-matter theories,''
JHEP {\bf 1602}, 093 (2016).
[arXiv:1512.00161 [hep-th]].
%%CITATION = WIS-12-15-NOV-DPPA%%
}

\lref\Rehren{
K.-H.~Rehren,
``Algebraic Conformal QFT'',
3rd Meeting of the French-Italian Research Team on Noncommutative Geometry and Quantum Physics Vietri sul Mare, 2009.
}

%\PaulyAMA
\lref\PaulyAMA{
  C.~Pauly,
  ``Strange duality revisited,''
Math.\ Res.\ Lett.\  {\bf 21}, 1353 (2014).
}

%\FidkowskiJUA
\lref\FidkowskiJUA{
  L.~Fidkowski, X.~Chen and A.~Vishwanath,
  ``Non-Abelian Topological Order on the Surface of a 3D Topological Superconductor from an Exactly Solved Model,''
Phys.\ Rev.\ X {\bf 3}, no. 4, 041016 (2013).
[arXiv:1305.5851 [cond-mat.str-el]].
%%CITATION = arXiv:1305.5851%%
}

\lref\VOstrik{
V.~Ostrik and M.~Sun,
``Level-Rank Duality Via Tensor Categories,''
Comm. Math. Phys. 326 (2014) 49-61.
[arXiv:1208.5131 [math-ph]].
}

%\DaviesNW
\lref\DaviesNW{
  N.~M.~Davies, T.~J.~Hollowood and V.~V.~Khoze,
  ``Monopoles, affine algebras and the gluino condensate,''
  J.\ Math.\ Phys.\  {\bf 44}, 3640 (2003).'
  [hep-th/0006011].%%CITATION = hep-th/0006011%%
}

%\KutasovVE
\lref\KutasovVE{
  D.~Kutasov,
  ``A Comment on duality in N=1 supersymmetric nonAbelian gauge theories,''
Phys.\ Lett.\ B {\bf 351}, 230 (1995).
[hep-th/9503086].
%%CITATION = hep-th/9503086%%
}

\lref\KinneyEJ{
  J.~Kinney, J.~M.~Maldacena, S.~Minwalla and S.~Raju,
  ``An Index for 4 dimensional super conformal theories,''
Commun.\ Math.\ Phys.\  {\bf 275}, 209 (2007).
[hep-th/0510251].
%%CITATION = hep-th/0510251%%
}

%\DaviesUW
\lref\DaviesUW{
  N.~M.~Davies, T.~J.~Hollowood, V.~V.~Khoze and M.~P.~Mattis,
  ``Gluino condensate and magnetic monopoles in supersymmetric gluodynamics,''
  Nucl.\ Phys.\ B {\bf 559}, 123 (1999).[hep-th/9905015].%%CITATION = hep-th/9905015%%
  }

%\LeeVP
\lref\LeeVP{
  K.~-M.~Lee and P.~Yi,
  ``Monopoles and instantons on partially compactified D-branes,''
  Phys.\ Rev.\ D {\bf 56}, 3711 (1997).
  [hep-th/9702107].%%CITATION = hep-th/9702107%%
  }
%\RadicevicYLA
\lref\RadicevicYLA{
  D.~Radicevic,
  ``Disorder Operators in Chern-Simons-Fermion Theories,''
JHEP {\bf 1603}, 131 (2016).
[arXiv:1511.01902 [hep-th]].
%%CITATION = SU-ITP-15-16%%
}

%\LeeVU
\lref\LeeVU{
  K.~-M.~Lee,
  ``Instantons and magnetic monopoles on R**3 x S**1 with arbitrary simple gauge groups,''
  Phys.\ Lett.\ B {\bf 426}, 323 (1998).[hep-th/9802012].%%CITATION = hep-th/9802012%%
  }

%\NiarchosAA
\lref\NiarchosAA{
  V.~Niarchos,
  ``R-charges, Chiral Rings and RG Flows in Supersymmetric Chern-Simons-Matter Theories,''
JHEP {\bf 0905}, 054 (2009).
[arXiv:0903.0435 [hep-th]].
%%CITATION = arXiv:0903.0435%%
}

%\IntriligatorER
\lref\IntriligatorER{
  K.~A.~Intriligator and N.~Seiberg,
  ``Phases of N=1 supersymmetric gauge theories and electric - magnetic triality,''
In *Los Angeles 1995, Future perspectives in string theory* 270-282.
[hep-th/9506084].
%%CITATION = hep-th/9506084%%
}
%\NiarchosJB
\lref\NiarchosJB{
  V.~Niarchos,
  ``Seiberg Duality in Chern-Simons Theories with Fundamental and Adjoint Matter,''
JHEP {\bf 0811}, 001 (2008).
[arXiv:0808.2771 [hep-th]].
%%CITATION = arXiv:0808.2771%%
}

%\BorokhovIB
\lref\BorokhovIB{
  V.~Borokhov, A.~Kapustin and X.~-k.~Wu,
  ``Topological disorder operators in three-dimensional conformal field theory,''
JHEP {\bf 0211}, 049 (2002).
[hep-th/0206054].
%%CITATION = hep-th/0206054%%
}

%\ParkWTA
\lref\ParkWTA{
  J.~Park and K.~-J.~Park,
  ``Seiberg-like Dualities for 3d N=2 Theories with SU(N) gauge group,''
[arXiv:1305.6280 [hep-th]].
%%CITATION = arXiv:1305.6280%%
}

%\BorokhovCG
\lref\BorokhovCG{
  V.~Borokhov, A.~Kapustin and X.~-k.~Wu,
  ``Monopole operators and mirror symmetry in three-dimensions,''
JHEP {\bf 0212}, 044 (2002).
[hep-th/0207074].
%%CITATION = hep-th/0207074%%
}

%\NakanishiHJ
\lref\NakanishiHJ{
  T.~Nakanishi and A.~Tsuchiya,
  ``Level rank duality of WZW models in conformal field theory,''
Commun.\ Math.\ Phys.\  {\bf 144}, 351 (1992).
%%CITATION = NU-MATH-002%%
}
\lref\NiarchosAH{
  V.~Niarchos,
  ``Seiberg dualities and the 3d/4d connection,''
JHEP {\bf 1207}, 075 (2012).
[arXiv:1205.2086 [hep-th]].
%%CITATION = arXiv:1205.2086%%
}

%\SeibergBZ
\lref\SeibergBZ{
  N.~Seiberg,
  ``Exact results on the space of vacua of four-dimensional SUSY gauge theories,''
Phys.\ Rev.\ D {\bf 49}, 6857 (1994).
[hep-th/9402044].
%%CITATION = hep-th/9402044%%
}

%\ZupnikRY
\lref\ZupnikRY{
   B.~M.~Zupnik and D.~G.~Pak,
   ``Topologically Massive Gauge Theories In Superspace,''
Sov.\ Phys.\ J.\  {\bf 31}, 962 (1988).
}

%\IvanovFN
\lref\IvanovFN{
   E.~A.~Ivanov,
   ``Chern-Simons matter systems with manifest N=2 supersymmetry,''
Phys.\ Lett.\ B {\bf 268}, 203 (1991).
}

%\NiemiRQ
\lref\NiemiRQ{
  A.~J.~Niemi and G.~W.~Semenoff,
  ``Axial Anomaly Induced Fermion Fractionization and Effective Gauge Theory Actions in Odd Dimensional Space-Times,''
Phys.\ Rev.\ Lett.\  {\bf 51}, 2077 (1983).
%%CITATION = Print-83-0988 (IAS,PRINCETON)%%
}

%\RedlichDV
\lref\RedlichDV{
  A.~N.~Redlich,
  ``Parity Violation and Gauge Noninvariance of the Effective Gauge Field Action in Three-Dimensions,''
Phys.\ Rev.\ D {\bf 29}, 2366 (1984).
%%CITATION = MIT-CTP-1128%%
}

\lref\AharonyGP{
  O.~Aharony,
  ``IR duality in d = 3 N=2 supersymmetric USp(2N(c)) and U(N(c)) gauge theories,''
Phys.\ Lett.\ B {\bf 404}, 71 (1997).
[hep-th/9703215].
%%CITATION = hep-th/9703215%%
}

%\AffleckAS
\lref\AffleckAS{
  I.~Affleck, J.~A.~Harvey and E.~Witten,
  ``Instantons and (Super)Symmetry Breaking in (2+1)-Dimensions,''
Nucl.\ Phys.\ B {\bf 206}, 413 (1982).
%%CITATION = PRINT-82-0478 (PRINCETON)%%
}

\lref\BeemMB{
  C.~Beem, T.~Dimofte and S.~Pasquetti,
  ``Holomorphic Blocks in Three Dimensions,''
[arXiv:1211.1986 [hep-th]].
%%CITATION = arXiv:1211.1986%%
}

\lref\HwangJH{
  C.~Hwang, H.~-C.~Kim and J.~Park,
  ``Factorization of the 3d superconformal index,''
[arXiv:1211.6023 [hep-th]].
%%CITATION = arXiv:1211.6023%%
}

\lref\KrattenthalerDA{
  C.~Krattenthaler, V.~P.~Spiridonov, G.~S.~Vartanov,
  ``Superconformal indices of three-dimensional theories related by mirror symmetry,''
JHEP {\bf 1106}, 008 (2011).
[arXiv:1103.4075 [hep-th]].
%%CITATION = arXiv:1103.4075%%
}

\lref\GaddeEN{
  A.~Gadde, L.~Rastelli, S.~S.~Razamat and W.~Yan,
  ``On the Superconformal Index of N=1 IR Fixed Points: A Holographic Check,''
JHEP {\bf 1103}, 041 (2011).
[arXiv:1011.5278 [hep-th]].
%%CITATION = arXiv:1011.5278%%
}

\lref\ImamuraWG{
  Y.~Imamura and D.~Yokoyama,
 ``N=2 supersymmetric theories on squashed three-sphere,''
Phys.\ Rev.\ D {\bf 85}, 025015 (2012).
[arXiv:1109.4734 [hep-th]].
%%CITATION = arXiv:1109.4734%%
}

%\deBoerKA
\lref\deBoerKA{
  J.~de Boer, K.~Hori, Y.~Oz and Z.~Yin,
  ``Branes and mirror symmetry in N=2 supersymmetric gauge theories in three-dimensions,''
Nucl.\ Phys.\ B {\bf 502}, 107 (1997).
[hep-th/9702154].
%%CITATION = hep-th/9702154%%
}

\lref\ClossetRU{
  C.~Closset, T.~T.~Dumitrescu, G.~Festuccia and Z.~Komargodski,
  ``Supersymmetric Field Theories on Three-Manifolds,''
JHEP {\bf 1305}, 017 (2013).
[arXiv:1212.3388 [hep-th]].
%%CITATION = PUPT-2432%%
}

\lref\ImamuraRQ{
  Y.~Imamura and D.~Yokoyama,
 ``$S^3/Z_n$ partition function and dualities,''
JHEP {\bf 1211}, 122 (2012).
[arXiv:1208.1404 [hep-th]].
%%CITATION = arXiv:1208.1404%%
}

\lref\KapustinSim{
A.~Kapustin,  2010 Simons Workshop talk, a video of this talk can be found at
{\tt
http://media.scgp.stonybrook.edu/video/video.php?f=20110810\_1\_qtp.mp4}
}

%\PolyakovFU
\lref\PolyakovFU{
  A.~M.~Polyakov,
  ``Quark Confinement and Topology of Gauge Groups,''
Nucl.\ Phys.\ B {\bf 120}, 429 (1977).
%%CITATION = NORDITA-76/33%%
}

\lref\newIS{
K.~Intriligator and N.~Seiberg,
  ``Aspects of 3d N=2 Chern-Simons-Matter Theories,''
[arXiv:1305.1633 [hep-th]].
%%CITATION = UCSD-PTH-12-17%%
}

\lref\BhattacharyaZY{
  J.~Bhattacharya, S.~Bhattacharyya, S.~Minwalla and S.~Raju,
  ``Indices for Superconformal Field Theories in 3,5 and 6 Dimensions,''
JHEP {\bf 0802}, 064 (2008).
[arXiv:0801.1435 [hep-th]].
%%CITATION = arXiv:0801.1435%%
}

%\JafferisUN
\lref\JafferisUN{
  D.~L.~Jafferis,
  ``The Exact Superconformal R-Symmetry Extremizes Z,''
JHEP {\bf 1205}, 159 (2012).
[arXiv:1012.3210 [hep-th]].
%%CITATION = arXiv:1012.3210%%
}

%\JafferisZI
\lref\JafferisZI{
  D.~L.~Jafferis, I.~R.~Klebanov, S.~S.~Pufu and B.~R.~Safdi,
  ``Towards the F-Theorem: N=2 Field Theories on the Three-Sphere,''
JHEP {\bf 1106}, 102 (2011).
[arXiv:1103.1181 [hep-th]].
%%CITATION = arXiv:1103.1181%%
}

%\IntriligatorID
\lref\IntriligatorID{
  K.~A.~Intriligator and N.~Seiberg,
  ``Duality, monopoles, dyons, confinement and oblique confinement in supersymmetric SO(N(c)) gauge theories,''
Nucl.\ Phys.\ B {\bf 444}, 125 (1995).
[hep-th/9503179].
%%CITATION = hep-th/9503179%%
}

%\SeibergPQ
\lref\SeibergPQ{
  N.~Seiberg,
  ``Electric - magnetic duality in supersymmetric nonAbelian gauge theories,''
Nucl.\ Phys.\ B {\bf 435}, 129 (1995).
[hep-th/9411149].
%%CITATION = hep-th/9411149%%
}

%\StrasslerFE
\lref\StrasslerFE{
  M.~J.~Strassler,
  ``Duality, phases, spinors and monopoles in $SO(N)$ and $spin(N)$ gauge theories,''
JHEP {\bf 9809}, 017 (1998).
[hep-th/9709081].
%%CITATION = hep-th/9709081%%
}
%\AharonyBX
\lref\AharonyBX{
  O.~Aharony, A.~Hanany, K.~A.~Intriligator, N.~Seiberg and M.~J.~Strassler,
  ``Aspects of N=2 supersymmetric gauge theories in three-dimensions,''
Nucl.\ Phys.\ B {\bf 499}, 67 (1997).
[hep-th/9703110].
%%CITATION = hep-th/9703110%%
}

%\IntriligatorNE
\lref\IntriligatorNE{
  K.~A.~Intriligator and P.~Pouliot,
  ``Exact superpotentials, quantum vacua and duality in supersymmetric SP(N(c)) gauge theories,''
Phys.\ Lett.\ B {\bf 353}, 471 (1995).
[hep-th/9505006].
%%CITATION = hep-th/9505006%%
}

%\IntriligatorAU
\lref\IntriligatorAU{
  K.~A.~Intriligator and N.~Seiberg,
  ``Lectures on supersymmetric gauge theories and electric - magnetic duality,''
Nucl.\ Phys.\ Proc.\ Suppl.\  {\bf 45BC}, 1 (1996).
[hep-th/9509066].
%%CITATION = hep-th/9509066%%
}

%\KarchUX
\lref\KarchUX{
  A.~Karch,
  ``Seiberg duality in three-dimensions,''
Phys.\ Lett.\ B {\bf 405}, 79 (1997).
[hep-th/9703172].
%%CITATION = hep-th/9703172%%
}
%\MaldacenaSS
\lref\MaldacenaSS{
  J.~M.~Maldacena, G.~W.~Moore and N.~Seiberg,
  ``D-brane charges in five-brane backgrounds,''
JHEP {\bf 0110}, 005 (2001).
[hep-th/0108152].
%%CITATION = hep-th/0108152%%
}

%\BanksZN
\lref\BanksZN{
  T.~Banks and N.~Seiberg,
  ``Symmetries and Strings in Field Theory and Gravity,''
Phys.\ Rev.\ D {\bf 83}, 084019 (2011).
[arXiv:1011.5120 [hep-th]].
%%CITATION = arXiv:1011.5120%%
}

%\GaiottoBE
\lref\GaiottoBE{
  D.~Gaiotto, G.~W.~Moore and A.~Neitzke,
  ``Framed BPS States,''
[arXiv:1006.0146 [hep-th]].
%%CITATION = arXiv:1006.0146%%
}
%\SafdiRE
\lref\SafdiRE{
  B.~R.~Safdi, I.~R.~Klebanov and J.~Lee,
  ``A Crack in the Conformal Window,''
[arXiv:1212.4502 [hep-th]].
%%CITATION = arXiv:1212.4502%%
}

%\KutasovNP
\lref\KutasovNP{
  D.~Kutasov and A.~Schwimmer,
  ``On duality in supersymmetric Yang-Mills theory,''
Phys.\ Lett.\ B {\bf 354}, 315 (1995).
[hep-th/9505004].
%%CITATION = hep-th/9505004%%
}

%\KutasovSS
\lref\KutasovSS{
  D.~Kutasov, A.~Schwimmer and N.~Seiberg,
  ``Chiral rings, singularity theory and electric - magnetic duality,''
Nucl.\ Phys.\ B {\bf 459}, 455 (1996).
[hep-th/9510222].
%%CITATION = hep-th/9510222%%
}

%\GiveonZN
\lref\GiveonZN{
  A.~Giveon and D.~Kutasov,
  ``Seiberg Duality in Chern-Simons Theory,''
Nucl.\ Phys.\ B {\bf 812}, 1 (2009).
[arXiv:0808.0360 [hep-th]].
%%CITATION = arXiv:0808.0360%%
}

\lref\HoriDK{
  K.~Hori and D.~Tong,
  ``Aspects of Non-Abelian Gauge Dynamics in Two-Dimensional N=(2,2) Theories,''
JHEP {\bf 0705}, 079 (2007).
[hep-th/0609032].
%%CITATION = hep-th/0609032%%
}

%\XuLXA
\lref\XuLXA{
  C.~Xu and Y.~Z.~You,
  ``Self-dual Quantum Electrodynamics as Boundary State of the three dimensional Bosonic Topological Insulator,''
Phys.\ Rev.\ B {\bf 92}, no. 22, 220416 (2015).
[arXiv:1510.06032 [cond-mat.str-el]].
%%CITATION = arXiv:1510.06032%%
}

\lref\NiarchosAH{
  V.~Niarchos,
  ``Seiberg dualities and the 3d/4d connection,''
JHEP {\bf 1207}, 075 (2012).
[arXiv:1205.2086 [hep-th]].
%%CITATION = arXiv:1205.2086%%
}

\lref\KapustinJM{
  A.~Kapustin and B.~Willett,
  ``Generalized Superconformal Index for Three Dimensional Field Theories,''
[arXiv:1106.2484 [hep-th]].
%%CITATION = arXiv:1106.2484%%
}

%\KimCMA
\lref\KimCMA{
  H.~Kim and J.~Park,
  ``Aharony Dualities for 3d Theories with Adjoint Matter,''
[arXiv:1302.3645 [hep-th]].
%%CITATION = arXiv:1302.3645%%
}

%\KapustinVZ
\lref\KapustinVZ{
  A.~Kapustin, H.~Kim and J.~Park,
  ``Dualities for 3d Theories with Tensor Matter,''
JHEP {\bf 1112}, 087 (2011).
[arXiv:1110.2547 [hep-th]].
%%CITATION = arXiv:1110.2547%%
}

\lref\AharonyGP{
  O.~Aharony,
  ``IR duality in d = 3 N=2 supersymmetric USp(2N(c)) and U(N(c)) gauge theories,''
Phys.\ Lett.\ B {\bf 404}, 71 (1997).
[hep-th/9703215].
%%CITATION = hep-th/9703215%%
}

%\WittenDS
\lref\WittenDS{
  E.~Witten,
  ``Supersymmetric index of three-dimensional gauge theory,''
In *Shifman, M.A. (ed.): The many faces of the superworld* 156-184.
[hep-th/9903005].
%%CITATION = hep-th/9903005%%
}

\lref\FestucciaWS{
  G.~Festuccia and N.~Seiberg,
  ``Rigid Supersymmetric Theories in Curved Superspace,''
JHEP {\bf 1106}, 114 (2011).
[arXiv:1105.0689 [hep-th]].
%%CITATION = arXiv:1105.0689%%
}

\lref\SpiridonovHF{
  V.~P.~Spiridonov and G.~S.~Vartanov,
  ``Elliptic hypergeometry of supersymmetric dualities II. Orthogonal groups, knots, and vortices,''
[arXiv:1107.5788 [hep-th]].
%%CITATION = arXiv:1107.5788%%
}
\lref\SpiridonovZR{
  V.~P.~Spiridonov and G.~S.~Vartanov,
  ``Superconformal indices for N = 1 theories with multiple duals,''
Nucl.\ Phys.\ B {\bf 824}, 192 (2010).
[arXiv:0811.1909 [hep-th]].
%%CITATION = arXiv:0811.1909%%
}

\lref\HoriPD{
  K.~Hori,
  ``Duality In Two-Dimensional (2,2) Supersymmetric Non-Abelian Gauge Theories,''
[arXiv:1104.2853 [hep-th]].
%%CITATION = arXiv:1104.2853%%
}

\lref\RomelsbergerEG{
  C.~Romelsberger,
  ``Counting chiral primaries in N = 1, d=4 superconformal field theories,''
Nucl.\ Phys.\ B {\bf 747}, 329 (2006).
[hep-th/0510060].
%%CITATION = hep-th/0510060%%
}

\lref\KapustinKZ{
  A.~Kapustin, B.~Willett and I.~Yaakov,
  ``Exact Results for Wilson Loops in Superconformal Chern-Simons Theories with Matter,''
JHEP {\bf 1003}, 089 (2010).
[arXiv:0909.4559 [hep-th]].
%%CITATION = arXiv:0909.4559%%
}

\lref\SpiridonovHF{
  V.~P.~Spiridonov and G.~S.~Vartanov,
  ``Elliptic hypergeometry of supersymmetric dualities II. Orthogonal groups, knots, and vortices,''
[arXiv:1107.5788 [hep-th]].
%%CITATION = AEI-2011-049%%
}

%\NaculichNC
\lref\NaculichNC{
  S.~G.~Naculich and H.~J.~Schnitzer,
  ``Level-rank duality of the U(N) WZW model, Chern-Simons theory, and 2-D qYM theory,''
JHEP {\bf 0706}, 023 (2007).
[hep-th/0703089 [HEP-TH]].
%%CITATION = hep-th/0703089%%
}

\lref\RazamatUV{
  S.~S.~Razamat,
  ``On a modular property of N=2 superconformal theories in four dimensions,''
JHEP {\bf 1210}, 191 (2012).
[arXiv:1208.5056 [hep-th]].
%%CITATION = arXiv:1208.5056%%
}
%\cite{Camperi:1990dk}
\lref\Camperi{
  M.~Camperi, F.~Levstein and G.~Zemba,
  ``The Large N Limit Of Chern-simons Gauge Theory,''
  Phys.\ Lett.\ B {\bf 247} (1990) 549.
  %%CITATION = PHLTA,B247,549;%%
}

%\cite{Mlawer:1990uv}
\lref\Mlawer{
  E.~J.~Mlawer, S.~G.~Naculich, H.~A.~Riggs and H.~J.~Schnitzer,
  ``Group level duality of WZW fusion coefficients and Chern-Simons link observables,''
  Nucl.\ Phys.\ B {\bf 352} (1991) 863.
  %%CITATION = NUPHA,B352,863;%%
}

%\DolanQI
\lref\DolanQI{
  F.~A.~Dolan and H.~Osborn,
  ``Applications of the Superconformal Index for Protected Operators and q-Hypergeometric Identities to N=1 Dual Theories,''
Nucl.\ Phys.\ B {\bf 818}, 137 (2009).
[arXiv:0801.4947 [hep-th]].
%%CITATION = arXiv:0801.4947%%
}

\lref\EagerHX{
  R.~Eager, J.~Schmude and Y.~Tachikawa,
  ``Superconformal Indices, Sasaki-Einstein Manifolds, and Cyclic Homologies,''
[arXiv:1207.0573 [hep-th]].
%%CITATION = arXiv:1207.0573%%
}

\lref\GaddeIA{
  A.~Gadde and W.~Yan,
  ``Reducing the 4d Index to the $S^3$ Partition Function,''
JHEP {\bf 1212}, 003 (2012).
[arXiv:1104.2592 [hep-th]].
%%CITATION = arXiv:1104.2592%%
}

\lref\DolanRP{
  F.~A.~H.~Dolan, V.~P.~Spiridonov and G.~S.~Vartanov,
  ``From 4d superconformal indices to 3d partition functions,''
Phys.\ Lett.\ B {\bf 704}, 234 (2011).
[arXiv:1104.1787 [hep-th]].
%%CITATION = arXiv:1104.1787%%
}

%\IntriligatorEX
\lref\IntriligatorEX{
  K.~A.~Intriligator and N.~Seiberg,
  ``Mirror symmetry in three-dimensional gauge theories,''
Phys.\ Lett.\ B {\bf 387}, 513 (1996).
[hep-th/9607207].
%%CITATION = hep-th/9607207%%
}

%\deBoerMP
\lref\deBoerMP{
  J.~de Boer, K.~Hori, H.~Ooguri and Y.~Oz,
  ``Mirror symmetry in three-dimensional gauge theories, quivers and D-branes,''
Nucl.\ Phys.\ B {\bf 493}, 101 (1997).
[hep-th/9611063].
%%CITATION = hep-th/9611063%%
}

\lref\ImamuraUW{
  Y.~Imamura,
 ``Relation between the 4d superconformal index and the $S^3$ partition function,''
JHEP {\bf 1109}, 133 (2011).
[arXiv:1104.4482 [hep-th]].
%%CITATION = arXiv:1104.4482%%
}

%\SeibergBZ
\lref\SeibergBZ{
  N.~Seiberg,
  ``Exact results on the space of vacua of four-dimensional SUSY gauge theories,''
Phys.\ Rev.\ D {\bf 49}, 6857 (1994).
[hep-th/9402044].
%%CITATION = hep-th/9402044%%
}

\lref\HamaEA{
  N.~Hama, K.~Hosomichi and S.~Lee,
  ``SUSY Gauge Theories on Squashed Three-Spheres,''
JHEP {\bf 1105}, 014 (2011).
[arXiv:1102.4716 [hep-th]].
%%CITATION = arXiv:1102.4716%%
}

%\AffleckAS
\lref\AffleckAS{
  I.~Affleck, J.~A.~Harvey and E.~Witten,
  ``Instantons and (Super)Symmetry Breaking in (2+1)-Dimensions,''
Nucl.\ Phys.\ B {\bf 206}, 413 (1982).
%%CITATION = PRINT-82-0478 (PRINCETON)%%
}

%\SeibergPQ
\lref\SeibergPQ{
  N.~Seiberg,
  ``Electric - magnetic duality in supersymmetric nonAbelian gauge theories,''
Nucl.\ Phys.\ B {\bf 435}, 129 (1995).
[hep-th/9411149].
%%CITATION = hep-th/9411149%%
}

%\CveticXN
\lref\CveticXN{
  M.~Cvetic, T.~W.~Grimm and D.~Klevers,
  ``Anomaly Cancellation And Abelian Gauge Symmetries In F-theory,''
JHEP {\bf 1302}, 101 (2013).
[arXiv:1210.6034 [hep-th]].
%%CITATION = arXiv:1210.6034%%
}

\lref\debult{
  F.~van~de~Bult,
  ``Hyperbolic Hypergeometric Functions,''
University of Amsterdam Ph.D. thesis
}

%\MoritaCS
\lref\MoritaCS{
  T.~Morita and V.~Niarchos,
  ``F-theorem, duality and SUSY breaking in one-adjoint Chern-Simons-Matter theories,''
Nucl.\ Phys.\ B {\bf 858}, 84 (2012).
[arXiv:1108.4963 [hep-th]].
%%CITATION = arXiv:1108.4963%%
}

\lref\Shamirthesis{
  I.~Shamir,
  ``Aspects of three dimensional Seiberg duality,''
  M. Sc. thesis submitted to the Weizmann Institute of Science, April 2010.
  }

\lref\slthreeZ{
  J.~Felder, A.~Varchenko,
  ``The elliptic gamma function and $SL(3,Z) \times Z^3$,'' $\;\;$
[arXiv:math/0001184].
}

\lref\SpiridonovZA{
  V.~P.~Spiridonov and G.~S.~Vartanov,
  ``Elliptic Hypergeometry of Supersymmetric Dualities,''
Commun.\ Math.\ Phys.\  {\bf 304}, 797 (2011).
[arXiv:0910.5944 [hep-th]].
%%CITATION = arXiv:0910.5944%%
}

\lref\ZwiebelWA{
  B.~I.~Zwiebel,
  ``Charging the Superconformal Index,''
JHEP {\bf 1201}, 116 (2012).
[arXiv:1111.1773 [hep-th]].
%%CITATION = arXiv:1111.1773%%
}

%\BeniniMF
\lref\BeniniMF{
  F.~Benini, C.~Closset and S.~Cremonesi,
  ``Comments on 3d Seiberg-like dualities,''
JHEP {\bf 1110}, 075 (2011).
[arXiv:1108.5373 [hep-th]].
%%CITATION = arXiv:1108.5373%%
}

%\ClossetVG
\lref\ClossetVG{
  C.~Closset, T.~T.~Dumitrescu, G.~Festuccia, Z.~Komargodski and N.~Seiberg,
  ``Contact Terms, Unitarity, and F-Maximization in Three-Dimensional Superconformal Theories,''
JHEP {\bf 1210}, 053 (2012).
[arXiv:1205.4142 [hep-th]].
%%CITATION = arXiv:1205.4142%%
}

%\ClossetVP
\lref\ClossetVP{
  C.~Closset, T.~T.~Dumitrescu, G.~Festuccia, Z.~Komargodski and N.~Seiberg,
  ``Comments on Chern-Simons Contact Terms in Three Dimensions,''
JHEP {\bf 1209}, 091 (2012).
[arXiv:1206.5218 [hep-th]].
%%CITATION = arXiv:1206.5218%%
}

\lref\SpiridonovHF{
  V.~P.~Spiridonov and G.~S.~Vartanov,
  ``Elliptic hypergeometry of supersymmetric dualities II. Orthogonal groups, knots, and vortices,''
[arXiv:1107.5788 [hep-th]].
%%CITATION = arXiv:1107.5788%%
}

%\ElitzurFH
\lref\ElitzurFH{
  S.~Elitzur, A.~Giveon and D.~Kutasov,
  ``Branes and N=1 duality in string theory,''
Phys.\ Lett.\ B {\bf 400}, 269 (1997).
[hep-th/9702014].
%%CITATION = hep-th/9702014%%
}

%\ElitzurHC
\lref\ElitzurHC{
  S.~Elitzur, A.~Giveon, D.~Kutasov, E.~Rabinovici and A.~Schwimmer,
  ``Brane dynamics and N=1 supersymmetric gauge theory,''
Nucl.\ Phys.\ B {\bf 505}, 202 (1997).
[hep-th/9704104].
%%CITATION = hep-th/9704104%%
}

%\KapustinGH
\lref\KapustinGH{
  A.~Kapustin,
  ``Seiberg-like duality in three dimensions for orthogonal gauge groups,''
[arXiv:1104.0466 [hep-th]].
%%CITATION = arXiv:1104.0466%%
}

%\HwangHT
\lref\HwangHT{
  C.~Hwang, K.~-J.~Park and J.~Park,
  ``Evidence for Aharony duality for orthogonal gauge groups,''
JHEP {\bf 1111}, 011 (2011).
[arXiv:1109.2828 [hep-th]].
%%CITATION = arXiv:1109.2828%%
}

\lref\SpiridonovWW{
  V.~P.~Spiridonov and G.~S.~Vartanov,
  ``Elliptic hypergeometric integrals and 't Hooft anomaly matching conditions,''
JHEP {\bf 1206}, 016 (2012).
[arXiv:1203.5677 [hep-th]].
%%CITATION = arXiv:1203.5677%%
}

\lref\DimoftePY{
  T.~Dimofte, D.~Gaiotto and S.~Gukov,
  ``3-Manifolds and 3d Indices,''
[arXiv:1112.5179 [hep-th]].
%%CITATION = arXiv:1112.5179%%
}

\lref\KimWB{
  S.~Kim,
  ``The Complete superconformal index for N=6 Chern-Simons theory,''
Nucl.\ Phys.\ B {\bf 821}, 241 (2009), [Erratum-ibid.\ B {\bf 864}, 884 (2012)].
[arXiv:0903.4172 [hep-th]].
%%CITATION = arXiv:0903.4172%%
}

%\WillettGP
\lref\WillettGP{
  B.~Willett and I.~Yaakov,
  ``N=2 Dualities and Z Extremization in Three Dimensions,''
[arXiv:1104.0487 [hep-th]].
%%CITATION = arXiv:1104.0487%%
}

\lref\KapustinXQ{
  A.~Kapustin, B.~Willett and I.~Yaakov,
  ``Nonperturbative Tests of Three-Dimensional Dualities,''
JHEP {\bf 1010}, 013 (2010).
[arXiv:1003.5694 [hep-th]].
%%CITATION = arXiv:1003.5694%%
}

\lref\ImamuraSU{
  Y.~Imamura and S.~Yokoyama,
  ``Index for three dimensional superconformal field theories with general R-charge assignments,''
JHEP {\bf 1104}, 007 (2011).
[arXiv:1101.0557 [hep-th]].
%%CITATION = arXiv:1101.0557%%
}

%\GaddeDDA
\lref\GaddeDDA{
  A.~Gadde and S.~Gukov,
  ``2d Index and Surface operators,''
[arXiv:1305.0266 [hep-th]].
%%CITATION = CALT-68.2932%%
}

\lref\HwangQT{
  C.~Hwang, H.~Kim, K.~-J.~Park and J.~Park,
  ``Index computation for 3d Chern-Simons matter theory: test of Seiberg-like duality,''
JHEP {\bf 1109}, 037 (2011).
[arXiv:1107.4942 [hep-th]].
%%CITATION = arXiv:1107.4942%%
}

\lref\GreenDA{
  D.~Green, Z.~Komargodski, N.~Seiberg, Y.~Tachikawa and B.~Wecht,
  ``Exactly Marginal Deformations and Global Symmetries,''
JHEP {\bf 1006}, 106 (2010).
[arXiv:1005.3546 [hep-th]].
%%CITATION = arXiv:1005.3546%%
}

%\IntriligatorID
\lref\IntriligatorID{
  K.~A.~Intriligator and N.~Seiberg,
  ``Duality, monopoles, dyons, confinement and oblique confinement in supersymmetric SO(N(c)) gauge theories,''
Nucl.\ Phys.\ B {\bf 444}, 125 (1995).
[hep-th/9503179].
%%CITATION = hep-th/9503179%%
}

  %\SeibergQD
\lref\SeibergQD{
  N.~Seiberg,
  ``Modifying the Sum Over Topological Sectors and Constraints on Supergravity,''
JHEP {\bf 1007}, 070 (2010).
[arXiv:1005.0002 [hep-th]].
%%CITATION = arXiv:1005.0002%%
}
%\BanksZN
\lref\BanksZN{
  T.~Banks and N.~Seiberg,
  ``Symmetries and Strings in Field Theory and Gravity,''
Phys.\ Rev.\ D {\bf 83}, 084019 (2011).
[arXiv:1011.5120 [hep-th]].
%%CITATION = arXiv:1011.5120%%
}

%\SeibergNZ
\lref\SeibergNZ{
  N.~Seiberg and E.~Witten,
  ``Gauge dynamics and compactification to three-dimensions,''
In *Saclay 1996, The mathematical beauty of physics* 333-366.
[hep-th/9607163].
%%CITATION = hep-th/9607163%%
}

%\AharonyCI
\lref\AharonyCI{
  O.~Aharony and I.~Shamir,
  ``On $O(N_c)$ d=3 N=2 supersymmetric QCD Theories,''
JHEP {\bf 1112}, 043 (2011).
[arXiv:1109.5081 [hep-th]].
%%CITATION = arXiv:1109.5081%%
}

%\GiveonSR
\lref\GiveonSR{
  A.~Giveon and D.~Kutasov,
  ``Brane dynamics and gauge theory,''
Rev.\ Mod.\ Phys.\  {\bf 71}, 983 (1999).
[hep-th/9802067].
%%CITATION = hep-th/9802067%%
}

%\KapustinPY
\lref\KapustinPY{
  A.~Kapustin,
  ``Wilson-'t Hooft operators in four-dimensional gauge theories and S-duality,''
Phys.\ Rev.\ D {\bf 74}, 025005 (2006).
[hep-th/0501015].
%%CITATION = hep-th/0501015%%
}

%\KapustinGH
\lref\KapustinGH{
  A.~Kapustin,
  ``Seiberg-like duality in three dimensions for orthogonal gauge groups,''
[arXiv:1104.0466 [hep-th]].
%%CITATION = arXiv:1104.0466%%
}

%\AharonyDHA
\lref\ARSW{
  O.~Aharony, S.~S.~Razamat, N.~Seiberg and B.~Willett,
  ``3d dualities from 4d dualities,''
[arXiv:1305.3924 [hep-th]].
%%CITATION = WIS-04-13-APR-DPPA%%
}

\lref\AST{
O.~Aharony, N.~Seiberg and Y.~Tachikawa,
  ``Reading between the lines of four-dimensional gauge theories,''
[arXiv:1305.0318 [hep-th]].
%%CITATION = WIS-03-13-APR-DPPA%%
}

\vskip-50pt
\Title{} {\vbox{\centerline{Level/rank Duality and}
\centerline{}
\centerline{Chern-Simons-Matter Theories}
 }}

\vskip-15pt
\centerline{Po-Shen Hsin${}^1$ and Nathan Seiberg${}^2$}
\vskip15pt
\centerline{\it ${}^1$ Department of Physics, Princeton University, Princeton, NJ 08544, USA}
\centerline{\it ${}^2$ School of Natural Sciences, Institute for Advanced Study, Princeton, NJ 08540, USA}
\vskip25pt

\noindent
We discuss in detail level/rank duality in three-dimensional Chern-Simons theories and various related dualities in three-dimensional Chern-Simons-matter theories.  We couple the dual Lagrangians to appropriate background fields (including gauge fields, spin$_c$ connections and the metric).  The non-trivial maps between the currents and the line operators in the dual theories is accounted for by mixing of these fields.  In order for the duality to be valid we must add finite counterterms depending on these background fields.  This analysis allows us to resolve a number of puzzles with these dualities, to provide derivations of some of them, and to find new consistency conditions and relations between them.  In addition, we find new level/rank dualities of topological Chern-Simons theories and new dualities of Chern-Simons-matter theories, including new boson/boson and fermion/fermion dualities.

\bigskip
\Date{August 2016}

%\draftmode

%%%%%%%%%%%%%%%%%%%%%%%%%%%%%%%%%%%%%%%%%%%%%%%%%%%%%%%%%%%%%%%%%%%%%%%%%%%%%%%%%%%%%%%%%%%%%%%%%%%%%%%%%%%%%%%%%%%%%%%%%%%%
\newsec{Introduction}

Level/rank duality is an interesting example of a surprising duality between two different theories based on two different set of variables.  It is sufficiently simple and tractable that it can be established rigorously, and it seems related to other dualities that are harder to analyze.

One of the simplest contexts where this duality arises is in the theory of $NK$ complex chiral fermions in two dimensions \NakanishiHJ\ (see also \refs{\XuNXA\Rehren\PaulyAMA-\VOstrik}). This leads to the equivalence of the chiral algebras
\eqn\SUSUlr{SU(N)_K \quad \longleftrightarrow\quad  {SU(NK)_1\over SU(K)_N}~,}
where the notation $G_L$ represents the chiral algebra of $G$ with level $L$ and in the right hand side of \SUSUlr\ we have a GKO coset \GoddardVK.  In the context of the corresponding three-dimensional Chern-Simons theory \SUSUlr\ is represented as the duality between theories based on the two Lagrangians
\eqn\SUSUlrL{\CL_{SU(N)_K} \quad \longleftrightarrow\  \quad \CL_{SU(NK)_1/ SU(K)_N}=\CL_{(SU(NK)_1\times SU(K)_{-N})/ \Z_K}~,}
where we used the Chern-Simons Lagrangian description of coset models of \MooreYH.  (In section 4.1 we will present another Lagrangian for this coset theory.)

Other versions of level/rank duality are
\eqn\SUUUU{\eqalign{
&SU(N)_{\pm K} \quad\longleftrightarrow\quad U(K)_{\mp N,\mp N}\cr
&U(N)_{K,K\pm N} \quad\longleftrightarrow\quad U(K)_{-N,-N\mp K}
~,}}
where $U(N)_{K,K+NK'}\equiv(SU(N)_K\times U(1)_{N(K+NK')})/\Z_N$ (see section 2) and  $N,K>0$.

Orientation-reversal exchanges the two dualities in the first line of \SUUUU.  The two dualities in the second line of \SUUUU\ are mapped to themselves under orientation-reversal (the left hand side and the right hand side of the dualities are exchanged).  Three of these dualities are well known (see e.g.\ the analysis of the Chern-Simons theory in \refs{\NaculichPA\MlawerUV-\NaculichNC} and in \DouglasEX\ and the discussion of the cohomology of the Grassmannian \WittenXI), but as far as we know, the fourth duality
\eqn\newduality{U(N)_{K,K- N} \quad\longleftrightarrow\quad U(K)_{-N,-N+ K}}
is new.

Although the dualities \SUUUU\ are common, they lead to several puzzles.
\item{1.} The two-dimensional central charges (and correspondingly the three-dimensional framing anomaly) do not match between the two sides of these dualities.  What should we add to them in order to fix it?
\item{2.} The chiral algebra of a standard two-dimensional rational conformal field theory includes only integral-dimension operators.  This is true in $SU(N)_K$ for all $K$, $U(N)_{K,K}$ for even $K$, and $ U(N)_{K,K\pm N}$ for odd $K$ (see section 2).  Correspondingly, the three-dimensional Chern-Simons descriptions of these theories do not depend on the choice of spin structure.  This is not true in the other theories in \SUUUU.  They are spin-Chern-Simons theories and depend on the choice of spin structure.\foot{Ordinary, non-spin Chern-Simons theories are independent of a choice of spin structure and the spin theories depend on that choice. The spin theories include a line with half-integral spin (hence it is nontrivial), which is transparent, i.e.\ it has no braiding with all the lines in the theory.  This line does not decouple from the rest of the theory, because it can be obtained by fusing other lines.  We can consider such a transparent line also in the non-spin theories, but it decouples from the rest of the theory.  It is intrinsically gravitational.  A useful example to keep in mind is $U(1)_K$ Chern-Simons theory with the Lagrangian ${K\over 4\pi}bdb$.  For even $K$ this theory is not spin and does not depend on the choice of spin structure.  For odd $K$ this is a spin theory.  Quantizing this theory with any $K$ on a Riemann surface leads to $K^h$ states, where $h$ is the genus.  But the expression for the number of independent line observables depends on $K\mod 2$.  For even $K$ there are $K$ lines $e^{in\oint b}$ with the line $e^{iK\oint b}$ being trivial.  For odd $K$ the line $e^{iK\oint b}$ is transparent and its spin is half-integral.  Therefore, it is non-trivial and the theory with odd $K$ has $2K$ distinct lines.  For more details see, e.g.\ \refs{\SeibergRSG,\SeibergGMD}.} How can a non-spin Chern-Simons theory be dual to a spin Chern-Simons theory?
\item{3.} Consider the cases where the two sides of the duality \SUUUU\ do not depend on the choice of spin structure, i.e.\ the first duality with even $N$ and the second duality with odd $N$ and $K$.  Even here, the two-dimensional conformal weights of the representations and correspondingly the three-dimensional spins of the line observables do not quite match between the two sides of the duality.  They match only modulo one-half.  This does not occur in \SUSUlrL.  But it is puzzling in \SUUUU.

\medskip

Below we will resolve these puzzles.  First, we will add to all the non-spin theories in \SUUUU\ an almost trivial, purely gravitational sector consisting of a single non-trivial line observable of spin one-half.  This sector was discussed in \SeibergGMD\ and will be reviewed in section 2.3.  (In the condensed matter literature this sector is often denoted as $\{1,f\}$.)  We can also try to add this sector to the spin theories in \SUUUU, but since they already include such a transparent line, this added sector does not change them.  After these additions all the theories in \SUUUU\ are spin theories.  We will show that after this modification of \SUUUU\ the second and third puzzle above are resolved.

Next, we will couple these theories to appropriate background fields.  These includes ordinary $U(1)$ gauge fields $B$, $C$, etc.\ a spin$_c$ connection $A$ (see section 2.1) and a metric $g$.  In order for the dualities \SUUUU\ to be valid with the background fields present one needs to take into account two facts.  First, the map between the background fields in the left hand side and the right hand side of the duality will be non-trivial.  This will account for a non-trivial map of the line observables.  (Because of this non-trivial map one cannot simply factor out the decoupled sector with the transparent line; the duality map mixes the two sectors.)  Second, as in \ClossetVP, we will have to add appropriate finite counterterms constructed out of the background fields.  These counterterms will ensure that the theories are mapped correctly and they will also account for the discrepancy in the framing anomaly mentioned in the first puzzle above.

Related dualities were found in $\CN=2$ supersymmetric theories in three dimensions \refs{\GiveonZN\BeniniMF\AharonyDHA-\ParkWTA}. These theories contain matter fields coupled to gauge fields with Chern-Simons terms.  There is enormous evidence supporting these dualities, but they cannot be rigorously derived.\foot{The new level/rank duality \newduality\ has not been embedded in this context.  But our discussion below can be repeated there and it leads to new supersymmetric dualities enjoying the same nontrivial tests.}  These theories contain both fermions and bosons and they clearly depend on the choice of spin structure.  Therefore, the second question above does not arise.  However, when the matter fields are given masses and are integrated out, these theories lead to the topological theories \SUUUU\ and we still need to add the appropriate, transparent, spin one-half line.

One approach to finding non-supersymmetric dualities is to study Chern-Simons theories coupled to matter in the fundamental representation with large $N$ and large $K$ with fixed $N/K$.  In some cases two different theories, one of them is fermionic and the other is bosonic were found \refs{\AharonyJZ,\GiombiKC,\AharonyNH} to be dual to the same gravitational Vasiliev theory (see e.g.\ \VasilievVF).  Another approach to finding such non-supersymmetric dualities with finite $N$ and $K$
is based on starting with a pair of dual supersymmetric theories and turning on a relevant operator that breaks supersymmetry.  If the flow to the infra-red is smooth, we should find a non-supersymmetric duality \refs{\JainGZA,\GurPCA}.  Motivated by this whole body work and the map between baryon operators and monopole operators in these theories \RadicevicYLA\ (see below) Aharony conjectured \AharonyMJS\ three boson/fermion dualities and adding to it a new duality, which is motivated by \newduality\ we will discuss:
\eqn\Oferdg{\eqalign{
N_f\ {\rm scalars\ with\ } SU(N)_K \quad &\longleftrightarrow \quad N_f\ {\rm  fermions\ with }\ U(K)_{-N+{N_f\over 2}} \cr
N_f\ {\rm  scalars\ with\ } U(N)_K\quad &\longleftrightarrow \quad N_f\ {\rm  fermions\ with\ } SU(K)_{-N+{N_f\over 2}}\cr
N_f\ {\rm  scalars\ with\ }U(N)_{K,K\pm N}\quad &\longleftrightarrow \quad N_f\ {\rm  fermions\ with\ } U(K)_{-N+{N_f\over 2}, -N\mp K+{N_f\over 2}}
.~ }}
Here the matter fields are in the fundamental representation of the gauge group and it is implicit that the scalars $\phi$ are at a $|\phi|^4$ fixed point.  In addition, we can reverse the orientation of our manifold (apply time-reversal) and obtain four other dualities, which differ from \Oferdg\ by an overall sign change of all the subscripts (the levels).  Finally, the $N=K=N_f=1$ version of these dualities were analyzed and coupled to appropriate background fields in \refs{\SeibergGMD,\KarchSXI,\MuruganZAL}, thus providing further evidence that they are right.

One of the common tests of a duality is that the theories in the two sides of a duality must have the same global symmetry.  Furthermore, the correlation functions of the currents of the global symmetry must be the same in the two dual theories.  These statements apply to ordinary continuous global symmetries and to Poincare symmetry, where the associated current is the energy momentum tensor.  A useful tool to study these correlation functions is based on coupling the system to classical background gauge fields for these symmetries.  This can be done also for discrete global symmetries. Then, the partition functions as a functionals of these background fields must match in the two dual theories.  This matching guarantees that the correlation functions of the currents both at separated points and at coincident points are the same in the two dual theories.

It is often the case that a naive way of coupling the dual theories to background fields does not lead to the same contact terms.  In that case, the proper map between the two dual theories involves also added appropriate counterterms in one side of the duality.  These counterterms should be local, well-defined functionals of the background fields.  In our case these will be Chern-Simons terms in the background fields \ClossetVP.  And as with all Chern-Simons terms, their coefficients must be properly quantized.

These added background fields with their proper counterterms also allow us find new dualities.  More explicitly, we denote a duality between theories described by the Lagrangians $\CL_1[B]$ and $\CL_2[B]$ that depend on the same background fields $B$ (and include appropriate counterterms) as
\eqn\LodLt{\CL_1[B] \quad \longleftrightarrow\quad \CL_2[B] ~.}
We can transform the Lagrangians to new Lagrangians that describe new theories
$\CL_1'[C]$ and $\CL_2'[C]$ and maintain the duality
\eqn\LodLtp{\CL_1'[C] \quad \longleftrightarrow\quad \CL_2'[C] ~.}
These operations include \WittenYA
\eqn\twoope{\eqalign{
&T\ :\ \CL[B] \quad \to\quad \CL'[B] =\CL[B] +{1\over 4\pi} BdB \cr
&S\ :\ \CL[B] \quad \to\quad \CL'[C] =\CL[b] -{1\over 2\pi} bdC  ~,}}
where $T$ adds a counterterm for the background field $B$ and $S$ promotes $B$ to a dynamical field $b$ and couples it to a new classical background field $C$.  These satisfy $S^2=(ST)^3=\CC$, where $\CC$ acts as charge conjugation $B\to -B$.
On a spin$_c$ manifold odd powers of $T$-operation require a spin$_c$ connection (more in section 2.1).
We can also perform orientation-reversal on the two sides of the duality and we can use other known dualities.  Performing a sequence of such operations on a given duality \LodLt\ we can generate many new dualities \LodLtp.

One of the main points of this note will be to couple the dualities \SUSUlr-\Oferdg\ to background fields with appropriate counterterms.  This will allow us to resolve the puzzles mentioned above.  It will also allow us to use the rigorously derived duality \SUSUlr, \SUSUlrL\ to derive a more precise version of the dualities \SUUUU, to relate the dualities \Oferdg, and to find new Chern-Simons matter dualities.

We should mention our notations.  We will use an equal sign between two Lagrangians $\CL_1=\CL_2$ when they are the same as functionals of the fields and we will use an arrow $\CL_1\ \longleftrightarrow\ \CL_2$ when as functionals they are different (they might even be functionals of a different number of fields), but the theories described by the two Lagrangians are dual; i.e.\ $\CL_1$ and $ \CL_2$ lead to the same physics.  All our background fields (except the metric $g$) will be denoted by upper case letters and all dynamical fields by lowercase letters. $A$ will denote a background spin$_c$ connection (see below) and all other gauge fields will be standard gauge fields.

In section 2 we will review and extend some preliminary background material.  We will review the notion of a spin$_c$ connection and some facts about $U(N)$ Chern-Simons theories and in particular the almost trivial theory $U(N)_1$.  We will also discuss $SU(N)_K$ and will show how to couple it to background fields.

Sections 3 - 6 will present various dualities with increasing level of complexity and decreasing level of an explicit map between them.

Section 3 will discuss a number of dualities, which can be derived by an explicit and elementary change of variables.  Section 4 will be devoted to the dualities \SUSUlr\ and \SUUUU.  The first of them is rigorously established, but there is no known explicit change of variables between the two dual theories.  Then, given \SUSUlr\ we will derive a more precise version of \SUUUU.

Section 5 addresses the conjectural boson/fermion dualities \Oferdg\ and section 6 will present several boson/boson and fermion/fermion dualities obtained from the conjectural boson/fermion dualities in section 5.  Among other things we will use our conjectured dualities to find a derivation of the self-duality of QED with two fermions \XuLXA.  That duality is closely related to the mirror symmetry of the $\CN=4$ supersymmetric version of that theory \IntriligatorEX\ and it has many generalizations and extensions including to certain $\CN=2$ dualities \AharonyBX.

For convenience we collect many of the equations we use often in an appendix.

%%%%%%%%%%%%%%%%%%%%%%%%%%%%%%%%%%%%%%%%%%%%%%%%%%%%%%%%%%%%%%%%%%%%%%%%%%%%%%%%%%%%%%%%%%%%%%%%%%%%%%%%%%%%%%%%%%%%%%%%%%%%
%%%%%%%%%%%%%%%%%%%%%%%%%%%%%%%%%%%%%%%%%%%%%%%%%%%%%%%%%%%%%%%%%%%%%%%%%%%%%%%%%%%%%%%%%%%%%%%%%%%%%%%%%%%%%%%%%%%%%%%%%%%%
%%%%%%%%%%%%%%%%%%%%%%%%%%%%%%%%%%%%%%%%%%%%%%%%%%%%%%%%%%%%%%%%%%%%%%%%%%%%%%%%%%%%%%%%%%%%%%%%%%%%%%%%%%%%%%%%%%%%%%%%%%%%
\newsec{Preliminaries}

%%%%%%%%%%%%%%%%%%%%%%%%%%%%
\subsec{A brief summary of spin$_c$ for Chern-Simons theory}

A spin$_c$-connection $A$ is characterized by
\eqn\defspinc{
\int_{\gamma_2} {dA\over 2\pi}={1\over 2}\int_{\gamma_2} w_2(M)\;{\rm mod}\;\Z,
}
for any 2-cycle $\gamma_2$. In particular $2A$ always has integral period and thus is a $U(1)$ gauge field.
For more details about the application of spin$_c$ connections to this problem, see e.g.\ \refs{\MetlitskiYQA,\SeibergRSG}.  There it was also related to the spin/charge relation of condensed matter physics.

We will use the gravitational Chern-Simons term
\eqn\omegagrav{
\int_{M=\partial X} {\rm CS_{grav}} = \pi\int_{X} \hat A(R)= {1\over 192\pi}\int_{X} \tr R\wedge R~,
}
where $X$ is a bulk four-manifold, whose boundary is our spacetime $M$.  For a given $M$ different non-spin $X$ can lead to different $\int_{M} {\rm CS_{grav}}$, which differ by $\pi\Z /8$ and therefore only $\exp(16i\int_M\CSg)$ is meaningful.

The framing anomaly of ${\rm CS_{grav}}$ will be important for matching the framing anomaly in the dualities we discussed later.
If the framing of spacetime is changed by $n$ units, $\int_{M} {\rm CS_{grav}}$ changes by $n\pi /24$.

Consider background $U(1)$ gauge fields $B,C$, a background spin$_c$-connection $A$, and a background metric.  Integrals over
linear combinations with integer coefficients of the terms of the following form are well-defined on a spin$_c$ manifold:
\eqn\rulespinc{\eqalign{
&{1\over 2\pi}  BdC \cr
&{1\over 4\pi}  BdB+{1\over 2\pi}  Bd A  \cr
& {\cal I}[g,A] \equiv{1\over 4\pi}  AdA + 2 \CSg \cr
&16 \CSg
}}
and therefore they are valid counterterms.\foot{In fact, all these terms can be expressed as linear combinations of ${\cal I}[g,A]$ with appropriate values of $A$ \SeibergRSG
\eqn\lind{\eqalign{
&{1\over 4\pi}  BdB+{1\over 2\pi}  Bd A= {\cal I}[g, A+B]- {\cal I}[g,A] + d(...)\cr
&{1\over 2\pi}BdC ={\cal I}[g,A+B+C] -{\cal I}[g,A+B] - {\cal I}[g,A+C]+{\cal I}[g,A] +d(...)\cr
&16\CSg = 9{\cal I}[g,A] -{\cal I}[g,3A] ~.}}
}
Finally, the relation to the  Atiyah-Patodi-Singer $\eta$-invariant is $\eta[g,A] = \left({1\over \pi} \int \CI[g,A]\right)\mod 2 \Z$.

Although we will often couple our theories to a background spin$_c$ connection $A$, in many cases the resulting theory will not depend on the details of $A$; we will be able to absorb a shift $A\to A'$ by a $U(1)$ field $A'-A$ in an appropriate shift of other dynamical or background fields.  The explicit coupling to $A$ will serve as a reminder that our theory does not depend on a choice of spin structure and will allow us to transform correctly line operators in dual theories (see below).

%%%%%%%%%%%%%%%%%%%%%%%%%%%%%%%%%%%%%%%%%%%%%%%%%%%%%%%%%%%%%%%%%%%%%%%%%%%%%%%%%%%%%%%%%%%%%%%%%%%%%%%%%%%%%%%%%%%%%%%%%%%%%%%%%%%%%%
%%%%%%%%%%%%%%%%%%%%%%%%%%%%%%%%%%%%%%%%%%%%%%%%%%%%%%%%%%%%%%%%%%%%%%%%%%%%%%%%%%%%%%%%%%%%%%%%%%%%%%%%%%%%%%%%%%%%%%%%%%%%%%%%%%%%%%
%%%%%%%%%%%%%%%%%%%%%%%%%%%%%%%%%%%%%%%%%%%%%%%%%%%%%%%%%%%%%%%%%%%%%%%%%%%%%%%%%%%%%%%%%%%%%%%%%%%%%%%%%%%%%%%%%%%%%%%%%%%%%%%%%%%%%%
\subsec{Some facts about $U(N)_{K,L}$}

Throughout this section the integer $N$ will be positive, but the integers $K$ and $K'$ can be positive or negative.

Define $U(N)_{K,L}$ as
\eqn\u{
U(N)_{K,K+NK'}= {SU(N)_K\times U(1)_{N(K+NK')} \over \Z_N} ~,
}
which is consistent only for $K,K'\in \Z$.
This is manifest when we write the corresponding Chern-Simons Lagrangian
\eqn\UNCS{{K\over 4\pi} \Tr \left[ bdb -{2i\over 3}b^3\right] +{K'\over 4\pi} (\Tr b) d(\Tr b)~,}
where $b$ is a $U(N)$ gauge field.
From here it is also clear that the $U(N)_{K,K+NK'}$ Chern-Simons theory depends on the spin structure, if and only if $K+K'$ is odd.

We will be mostly interested in $U(N)_{K,K}$ ($K'=0$), which we will abbreviate as $U(N)_{K}$ and $U(N)_{K,K\pm N}$ ($K'=\pm 1$).

Next, we couple these theories to background fields in such a way that they do not depend on the spin structure.  Specifically, we couple the non-spin theories to a background $U(1)$ gauge field $C$ and the spin theories to a background spin$_c$-connection $A$.
For $U(N)_{K}$ we have
\eqn\UKKCS{\eqalign{
&\CL_{U(N)_{K}} [b;C]
\equiv {K\over 4\pi} \Tr \left[bdb-{2i\over 3}b^3\right]
+{1\over 2\pi} (\Tr b) dC  \quad\quad  {\rm for\ even}\ K\cr
&\CL_{U(N)_{K}} [b;A]
\equiv {K\over 4\pi} \Tr \left[bdb-{2i\over 3}b^3\right]
+{1\over 2\pi} (\Tr b) dA \quad\quad  {\rm for\ odd}\ K}~.}
These can be combined as
\eqn\UKKCSa{
\CL_{U(N)_{K}} [b;C+KA]
\equiv {K\over 4\pi} \Tr \left[bdb-{2i\over 3}b^3\right]
+{1\over 2\pi} (\Tr b) d\left(C+KA\right),
}
where the background field $C+KA$ is $U(1)$ for even $K$ and a spin$_c$ connection for odd $K$. It is important to stress the difference between the two expressions in \UKKCS, which is not manifest in \UKKCSa.  The even $K$ theory in \UKKCS\ is non-spin, while the odd $K$ theory is a spin theory.  Correspondingly, the odd $K$ theory has a transparent line observable with spin one-half.  Below we will also turn the even $K$ theory into a spin theory by adding to it a decoupled transparent line with spin one-half.

An important example is the special case of $N=1$
\eqn\UKKCSao{
\CL_{U(1)_{K}} [b;C+KA]
\equiv {K\over 4\pi} bdb
+{1\over 2\pi} b d\left(C+KA\right) ~.
}

It will be useful below that changing the orientation is the same as $K\to -K$ with $C\to -C$, $A\to A$.  (If we view $A$ as the electromagnetic field, then our definition of orientation change is what is normally called $\CC\CT$; i.e.\ time-reversal combined with charge conjugation.)

Similarly, for $U(N)_{K,K\pm N}$
\eqn\UKNplusKCS{\eqalign{
&\CL_{U(N)_{K,K\pm N}}[b;C]\equiv
{K\over 4\pi} \Tr \left[bdb-{2i\over 3}b^3\right]\pm{1 \over 4\pi} (\Tr b) d(\Tr b)
+{1\over 2\pi} (\Tr b)dC \quad  {\rm for\ odd}\ K\cr
&\CL_{U(N)_{K,K\pm N}}[b;A]\equiv
{K\over 4\pi} \Tr \left[bdb-{2i\over 3}b^3\right]\pm{1 \over 4\pi} (\Tr b) d(\Tr b)
+{1\over 2\pi} (\Tr b)d A\quad  {\rm for\ even}\ K~.
}}
These can be combined as
\eqn\UKNplusKCSa{\eqalign{
\CL_{U(N)_{K,K\pm N}}[b;C+(K\pm 1)A]\equiv&
{K\over 4\pi} \Tr \left[bdb-{2i\over 3}b^3\right]\pm{1 \over 4\pi} (\Tr b) d(\Tr b)\cr
&+{1\over 2\pi} (\Tr b)d\left( C+(K\pm 1)A\right)~,
}}
where the background field $C+(K\pm 1)A$ is spin$_c$ for even $K$ and $U(1)$ for odd $K$.
In the special case $N=1$ this coincides with \UKKCSao\ with $K\to K\pm 1$.  Note that \UKKCS-\UKNplusKCSa\ are consistent with \rulespinc.

Even though the Lagrangians \UKKCSa,\UKNplusKCSa\ depend on two background fields $A$ and $C$, in fact the dependence is only on a certain linear combination of them.  Therefore, for $K$ even in \UKKCSa\ and $K$ odd in \UKNplusKCSa\ the Lagrangians depend only on an even multiple of $A$ and as such the dependence on $A$ can be absorbed in a redefinition of $C$. For $K$ odd in \UKKCSa\ and $K$ even in \UKNplusKCSa\ $A$ cannot be removed.  But the dependence on the specific choice of $A$ can be absorbed in $C$; the difference between the choices $A$ and $A'$ can be absorbed in a shift of $C$ by a multiple of $A-A'$, which is an ordinary $U(1)$ field.  (See the comment at the end of section 2.1.)

We can define monopole operators in the Chern-Simons theory by removing a point from  spacetime and specifying the flux through a sphere around it.  Because of the equations of motion that flux must be localized at points on the sphere, and we will take it to be at a single point.  Therefore, the monopole operator can be thought of as creating a line with a trivial holonomy around it.  In $U(N)_K$ with even $K$ this line has integer spin and therefore it is completely trivial.  We can say that this monopole operator is a trivial bosonic operator; i.e.\ it is like the identity operator.  In $U(N)_K$ with odd $K$ this operator is fermionic and it creates a transparent line with half-integral spin.  This reflects the fact that the even $K$ theory is non-spin and the odd $K$ theory is a spin Chern-Simons theory.  Correspondingly, the even $K$ theory is coupled to a $U(1)$ background field $C$ and the odd $K$ theory is coupled to a spin$_c$ connection $A$.  The situation with $U(N)_{K,K\pm N}$ is the opposite.  Therefore the coupling to the background fields is also reversed, as in \UKNplusKCS.  Below we will couple these Chern-Simons theories to matter fields and then the monopole operators will be less trivial.

%%%%%%%%%%%%%%%%%%%%%%%%%%%%%%%%%%%%%%%%%%%%%%%%%%%%%%%%%%%%%%%%%%%%%%%%%%%%%%%%%%%%%%%%%%%%%%%%
\subsec{An almost trivial theory $U(N)_1$}

This is a slight generalization of Appendix B in \SeibergGMD, where more details can be found.

The special case $U(N)_{1}$ is almost trivial.  When quantizing this topological quantum field theory on a Riemann surface the Hilbert space consists of a single state and the partition function on any three manifold is a phase.  So this theory is dual to a classical field theory
 \eqn\UNonei{\CL_{U(N)_1}[A] \quad  \longleftrightarrow \quad - N{\cal I}[g,A] =-{N\over 4\pi} AdA - 2N{\rm CS_{grav}}~,}
where the arrow means that the two Lagrangians describe dual theories. This expression makes manifest the framing anomaly of the theory.  When the theory is quantized on a manifold with a boundary anomaly considerations force us to place degrees of freedom on the boundary.  One choice is placing $N$ chiral Dirac fermions there.

These properties of the theory can be used to show that
\eqn\Uspec{
\CL_{U(M)_{1}} [A] + \CL_{U(N)_{1}}[A] \quad  \longleftrightarrow \quad
\CL_{U(M+N)_{1}}[A] ~,}
where we suppressed the dynamical gauge fields.  We can extend the range of $N$ and preserve this property by defining
\eqn\triva{\eqalign{
&\CL_{U(-N)_{1}}[A]\equiv \CL_{U(N)_{-1}}[A] \cr
&\CL_{U(0)_{1}}[x,y;A]\equiv\CL_0[x,y;A]\equiv
{1\over 4\pi} xdx+{1\over 2\pi} xd(y+A)~. }}
Below we will often use the Lagrangian $\CL_0$, but we will not use $U(N)$ with negative $N$.

Even though these theories are almost trivial, they do have a non-trivial line observable.  The Wilson line in the fundamental representation of $U(N)$ is
a transparent line with spin one-half.

Because of \UNonei\ the theory based on the Lagrangian \triva\ is dual to a theory with vanishing Lagrangian.  Yet, we will use $\CL_0[A]$ as a Lagrangian to denote the fact that the theory includes a transparent line with spin one-half.

\subsec{A useful fact}

We will often use the fact that the Chern-Simons theory based on
\eqn\triCSt{\CL_{cv}={1\over 2\pi} c dv}
(with $c$ and $v$ $U(1)$ gauge fields) is trivial \WittenYA.\foot{This is the special case of $N=1$, $K=0$ of the Lagrangian for  a $\Z_N$ gauge theory of level $K$,  whose Lagrangian is \refs{\MaldacenaSS,\BanksZN,\KapustinGUA}
\eqn\ZNL{\eqalign{
&\CL_{\left(\Z_N\right)_K}[x,y;A]= {K\over 4\pi} xdx+{N\over 2\pi} xdy +{1\over 2\pi } x dA  \quad  {\rm for} \quad  K\ {\rm odd}\cr
&\CL_{\left(\Z_N\right)_K}[x,y;B]= {K\over 4\pi} xdx+{N\over 2\pi} xdy +{1\over 2\pi } x dB  \quad  {\rm for} \quad  K\ {\rm even} ~,
}}
where the theories with even $K$ are non-spin and the theories with odd $K$ are spin.}
Specifically, in many cases we will encounter a Lagrangian like
\eqn\eqnwitL{\CL(v,...) + {1\over 2\pi} c dv}
with $\CL$ independent of $c$.  Then, if the theory is non-spin, $c$ acts as a Lagrange multiplier setting $v$ to a pure gauge, which can be set to zero in $\CL$.  For example, when $\CL={L\over 2\pi } vdv + {1\over 2\pi} v de $ (with $e$ a linear combination with integer coefficients of gauge fields), the fact that the integral over $c$ sets $v$ to zero can be seen explicitly by shifting $c\to c-Lv -e$.

If $\CL$ describes a spin theory, the situation is not that simple.  For example, let $\CL={1\over 4\pi} vdv + {1\over 2\pi}vdA$, which is consistent with \rulespinc.  Now we cannot shift $c$ in \eqnwitL\ to eliminate the ${1\over 4\pi }vdv$ term.  Indeed, the theory $\CL +{1\over 2\pi} cdv$ is precisely the theory \triva, which has a line with spin a half.  However, remembering that the theory is a spin theory, we can still set $v$ to zero, but we should remember that the resulting theory should include that spin a half line.  One way to see that is to add to $\CL$ the theory $\CL_0$ \triva\ and replace \eqnwitL\ with
\eqn\eqnwitLs{\CL[v,c,x,y;A]={1\over 4\pi} vdv + {1\over 2\pi}vdA + {1\over 2\pi} c dv+{1\over 4\pi} xdx +{1\over 2\pi} xd(y+A) ~.}
As we explained above, this does not change the theory.  Now, we can change variables
\eqn\eqnitLsr{\CL[v,c+y,x-v, y+v;A]= {1\over 2\pi} c dv+{1\over 4\pi} xdx +{1\over 2\pi} xd(y+A)}
and see that we find the same answer as with $v$ to zero.
This can be generalized to general spin theory.

%%%%%%%%%%%%%%%%%%%%%%%%%%%%%%%%%%%%%%%%%%%%%%%%%%%%%%%%%%%%%%%%%%%%%%%%%%%%%%%%%%%%%%%%%%%%%%%%%%%%%%%
\subsec{$SU(N)_K$}

Using our $U(N)$ theories \UKKCS,\UKNplusKCS\ with background $U(1)$ field $C$ we can find the Lagrangians for $SU(N)$ theories by promoting $C$ to a dynamical field, which we will denote by $c$.  These impose the constraint that $\Tr b$ is trivial.  We can also use this Lagrangians to couple the $SU(N)$ theory to new $U(1)$ background fields $B$
\eqn\SU{\eqalign{
&\CL_{SU(N)_K}[b,c;B]
\equiv{K\over 4\pi} \Tr \left[bdb-{2i\over 3}b^3\right]
+{\epsilon_K \over 4\pi}(\Tr b)d(\Tr b)
+{1\over 2\pi} cd\left(\Tr b+B\right) \cr
&\epsilon_K=\cases{0& $K$ even \cr 1& $K$ odd and positive \cr -1 & $K$ odd and negative~.
}}}

This Lagrangian remains consistent, if we shift the definition of $\epsilon_K$ by an even (possibly $K$ dependent) integer $2n$.  But then we can redefine $c \to c - n(\Tr b -B)$ to shift $\epsilon_K$ back to \SU\ and find instead a counterterm ${n\over 2\pi}BdB$. So the only freedom in the definition \SU\ is in the choice of this counterterm.\foot{Here we used  \UKKCS,\UKNplusKCS\ and turned the background $C$ into a dynamical field.  What would have happened if we did the same thing starting with the equations with background $A$? Gauging it amounts to replacing $A\to c+A$ with $c$ a dynamical $U(1)$ gauge field:
\eqn\SUdz{
{K\over 4\pi} \Tr \left[bdb-{2i\over 3}b^3\right]
+{\epsilon_K+1\over 4\pi}(\Tr b)d(\Tr b)
+{1\over 2\pi} cd( \Tr b +B) +{1\over 2\pi} (\Tr b) d A  ~.
}
Unlike \SU\ the theory based on \SUdz\ is a spin theory.  Therefore, we can follow section 2.4 and as in the discussion around \eqnwitLs,\eqnitLsr\ we can add the Lagrangian $\CL_0$ \triva\ without affecting the outcome and use a change of variables like \eqnitLsr\ (with an appropriate shift of $c$ by $B$) to find that \SUdz\ plus $\CL_0$ is the same as
\eqn\SUgA{\CL_{SU(N)_K}[b,c;B] + \CL_0 [x,y;A]-{1\over 2\pi} AdB + {1\over 4\pi} BdB ~.}
Therefore, gauging $A$ in \UKKCS,\UKNplusKCS\ leads to the same theories as gauging $C$ except that these are spin theories and we have to add some counterterms.
}  It is important to stress that $c$ implements the constraint $\Tr b=-B$, but we cannot simply replace the term with $\epsilon_K$ in \SU\ by ${\epsilon_K\over 4\pi} BdB$.  (See section 2.4.)

Note that with the definition \SU\ the theory with $K$ is related to the theory with $-K$ by a change of orientation combined with $c\to -c$ (without transforming $b$ and $B$).

For trivial $B$ the Lagrangian \SU\ describes the standard non-spin $SU(N)_K$ Chern-Simons theory.  But the coupling of $B$ leads to additional possibilities.  If $B$ is topologically nontrivial, then $\Tr b$ is nontrivial and correspondingly, the functional integral can receive its contribution from a nontrivial $SU(N)/\Z_N$ bundle.  Note that in an $SU(N)/\Z_N$ theory we sum over such nontrivial bundles.  Here, as our theory is an $SU(N)$ gauge theory, we do not sum over such bundles.  Instead, depending on $B$ we have a given such bundle.  We see that, as emphasized in \KapustinGUA, $SU(N)/\Z_N$ bundles can appear as observables in the $SU(N)$ theory.  And here we control them using the background field $B$ in \SU.

We will find it useful to define
\eqn\SUh{\eqalign{
\hat \CL_{SU(N)_K}&[b,c,x, y;B]\cr
&\equiv{ K\over 4\pi} \Tr \left[bdb-{2i\over 3}b^3\right]
+{1\over 2\pi} (c+KA)d( \Tr b +B)  +\CL_0[x,y;A]+{\epsilon_K\over 4\pi }BdB~,}}
where $\CL_0[A]$ is the almost trivial theory \triva.
Note that it is consistent with \rulespinc.  For even $K$ it is trivially the same as $\CL_{SU(N)_K}[B]+\CL_0[A]$ (simply shift $c\to c-KA$).  For odd $K$ this is not true, but these two theories are still related.  To see that, follow the discussion around  \eqnitLsr\ and write
\eqn\SUddu{\eqalign{
\hat \CL_{SU(N)_K}[b,c,x, y;B] = \CL_{SU(N)_K}&[b,c+ \half (K^2-\epsilon_K)\Tr b+ \half (K^2+\epsilon_K)B -Ky;B]\cr
 &+\CL_0[x+K(\Tr b+B), y-K(\Tr b+B);A] ~,}}
where the change of variables is valid because $K^2\pm\epsilon_K$ are even.
This can be written as the duality
\eqn\SUdduu{
  \hat \CL_{SU(N)_K}[B] \quad\longleftrightarrow\quad \CL_{SU(N)_K}[B]+\CL_0[A]~,}
where we suppressed the dynamical fields.

We will often extend \SUh\ and use the definitions
\eqn\SUhg{\eqalign{
&\hat\CL_{SU(N)_K}[B]\quad\longleftrightarrow\quad \CL_{SU(N)_K}[B] +\CL_0[A] \cr
&\hat\CL_{U(N)_K}[B+KA]\equiv \CL_{U(N)_K}[B+KA] +\CL_0[A] \cr
&\hat\CL_{U(N)_{K,K\pm N}}[B+(K\pm 1)A]\equiv \CL_{U(N)_{K,K\pm N}}[B+(K\pm 1)A] +\CL_0[A]  ~.}}
As we said above, $\CL_0[A]$ is dual to a theory with a vanishing Lagrangian, but we will use it to denote the fact that it has a transparent line with spin one-half.  Therefore, our notation is such that there is a difference between the theory based on $\CL_{SU(N)_K}[B]$, which is not spin and the theory based on $\hat \CL_{SU(N)_K}[B]$, which is spin.  The latter theory has twice as many line observables because it includes also the transparent line of $\CL_0[A]$.  We should remind the reader that even though the presence of $\CL_0[A]$ leads to the transparent line, the theory still does not depend on the choice of spin structure.

For the $U(N)$ theories in \SUhg\ the situation is different.  For $K$ odd $U(N)_K$ is a spin theory and correspondingly, $B+KA$ is a spin$_c$ connection and $\CL_{U(N)_K}[B+KA]$ already includes the transparent line.  Therefore, in this case $\hat\CL_{U(N)_K}[B+KA]\ \longleftrightarrow\ \CL_{U(N)_K}[B+KA]$.  Similarly, for $K$ even $\hat\CL_{U(N)_{K,K\pm N}}[B+(K\pm 1)A]\ \longleftrightarrow\ \CL_{U(N)_{K,K\pm N}}[B+(K\pm 1)A]$.
%%%%%%%%%%%%%%%%%%%%%%%%%%%%%%%%%%%%%%%%%%%%%%%%%%%%%%%%%%%%%%%%%%%%%%%%%%%%%%%%%%%%%%%%%%%

%%%%%%%%%%%%%%%%%%%%%%%%%%%%%%%%%%%%%%%%%%%%%%%%%%%%%%%%%%%%%%%%%%%%%%%%%%%%%%%%%%%%%%%%%%%%%%%%%%%%%%%%%%%%%%
%%%%%%%%%%%%%%%%%%%%%%%%%%%%%%%%%%%%%%%%%%%%%%%%%%%%%%%%%%%%%%%%%%%%%%%%%%%%%%%%%%%%%%%%%%%%%%%%%%%%%%%%%%%%%%
%%%%%%%%%%%%%%%%%%%%%%%%%%%%%%%%%%%%%%%%%%%%%%%%%%%%%%%%%%%%%%%%%%%%%%%%%%%%%%%%%%%%%%%%%%%%%%%%%%%%%%%%%%%%%%
\newsec{Simple explicit examples of level/rank duality}

Here we discuss three simple dualities, which can be derived by a straightforward change of variables.  They demonstrate several points, which will be essential below.

\subsec{A simple Abelian example $U(1)_2 \longleftrightarrow  U(1)_{-2}$}

We start with
\eqn\UtwoUo{\CL_{U(1)_2}[b;B] +\CL_{U(1)_{-1}}[c;A] = {2 \over 4\pi} bdb +{1\over 2\pi} bdB - {1 \over 4\pi} cdc +{1\over 2\pi} cdA ~.}
A simple change of variables leads to
\eqn\UtwoUoc{\eqalign{
\CL_{U(1)_2}[b+c-B;B] +&\CL_{U(1)_{-1}}[c+2b-B;A] \cr
=& \CL_{U(1)_{-2}}[b;B+2A] +\CL_{U(1)_{1}}[c;A] -{1\over 4\pi}BdB-{1\over 2\pi} BdA~,}}
thus establishing the duality
\eqn\dualto{U(1)_2 \times U(1)_{-1}\quad  \longleftrightarrow \quad  U(1)_{-2} \times U(1)_1 ~,}
which we can write as
\eqn\dualtol{\CL_{U(1)_2}[B] +\CL_{U(1)_{-1}}[A]\quad \longleftrightarrow \quad  \CL_{U(1)_{-2}}[B+2A] +\CL_{U(1)_{1}}[A]-{1\over 4\pi}BdB-{1\over 2\pi} BdA ~,}
or, using \UNonei\ and the definition \SUhg, as
\eqn\dualtola{\hat\CL_{U(1)_2}[B] \quad  \longleftrightarrow \quad   \hat \CL_{U(1)_{-2}}[B+2A] -{1\over 4\pi}BdB-{1\over 2\pi} BdA -2{\cal I}[g,A]  ~,}
where we used $\hat\CL_{U(1)_{\pm 2}}$ to remind ourselves that the theory includes a transparent line with spin one-half.  As we commented at the end of section 2.1, neither side of the duality depends on the details of $A$; changing $A\to A'$ can be absorbed in a redefinition of the dynamical fields. This duality was derived in \SeibergRSG\ and was used in \SeibergGMD.  It had appeared earlier without the background fields and the essential counterterms in \FidkowskiJUA, where the theory of $\hat \CL_{U(1)_2}[B] $ was denoted as $U(1)_2\times \{1,f\}$ and was referred to as the semion/fermion theory.

We would like to make several comments about this duality, which will be useful below.
\item{1.} In the change of variables \UtwoUoc\ the dynamical fields $b$ and $c$ transform as a valid linear combination of the dynamical fields and the background fields $B$ and $A$.  This change of variables preserves the fact that they are $U(1)$ fields and therefore the coefficient of $A$ must be even (it vanishes).
\item{2.} The background fields $B$ and $A$ also transform, but their transformations are independent of the dynamical fields.  Also, they remain $U(1)$ and spin$_c$ fields respectively.
\item{3.} The duality involves a nontrivial counterterm of the background fields.  These terms are consistent with \rulespinc.
\item{4.} The form of the duality \dualtola\ might suggest to write it, up to counterterms, as $U(1)_2 \longleftrightarrow U(1)_{-2}$.  Below we will often use such shorthand notation, but it should be emphasized that it is misleading.  This duality is valid only when we include a transparent spin one-half line on both sides of the duality.  It couples to $A$ with charge one.  This is manifest in \dualtol, \dualtola.
\item{5.} In the left hand side of \dualtola\ the background field $B$ couples only to $b$ and the background field $A$ couples only to the transparent line.  This is not true in the right hand side of the duality, where the dynamical $U(1)_{-2}$, $\tilde b$ field couples to $B+2A$.  Explicitly, the line  $e^{i\oint b}$ of $U(1)_2$ carries charge $-\half$ under $B$ and is neutral under $A$.  (This follows from the equation of motion $2db+dB=0$.)   This should be contrasted with the basic line in the right hand side of the duality
    \eqn\maplt{e^{i\oint \tilde b} = e^{i \oint (-b+c)} ~,}
    which carries charge $\half$ under $B$ and charge $1$ under $A$.  This reflects the change of variables in \UtwoUoc\ and the map \maplt.  Such a nontrivial mapping of the lines will be common below and it is accounted for by the nontrivial map of the background fields.
\item{6.} It is instructive to see how orientation-reversal (or time-reversal) acts on \dualtola.  We let it act on the background fields as $A\to A$ and $B\to -B-2A$ and then \dualtola\ is mapped to itself (after moving the counterterms to the left hand side).

\subsec{A simple non-Abelian example $SU(N)_1 \longleftrightarrow  U(1)_{-N}$}

The second simple duality involves non-Abelian fields.  Consider \SUh\ for $K=1$
\eqn\SUhda{\eqalign{
\hat \CL_{SU(N)_1}&[b,c,x, y;B]\cr
=&{ 1\over 4\pi} \Tr \left[bdb-{2i\over 3}b^3\right]
+{1\over 2\pi} (c+A)d( \Tr b +B)+{1\over 4\pi }BdB+\CL_0[x,y;A]\cr
= &\CL_{U(N)_1}[b+c\unit;A] +\CL_{U(1)_{-N}}[c;B-NA]+{1\over 2\pi} AdB +{1\over 4\pi} BdB+\CL_0 [x,y;A]~,}}
which can be written, using \UNonei, as
\eqn\simpledua{\hat \CL_{SU(N)_1}[B] \quad  \longleftrightarrow \quad  \hat \CL_{U(1)_{-N}}[B-NA]+{1\over 2\pi} AdB +{1\over 4\pi} BdB-N{\cal I}[g,A] ~.}

All the comments we made after \dualtola\ are clearly true also here. (In fact, for $N=2$ the known relation $SU(2)_1\longleftrightarrow U(1)_2$ shows that the dualities \dualtola\ and \simpledua\ are the same.)  In particular, the presence of a transparent spin one-half line and the dependence on $A$ are essential.  Below we will often abbreviate this duality in the misleading form $SU(N)_1  \longleftrightarrow U(1)_{-N}$, but we should remember to keep in mind these subtleties.  In order to remind ourselves about that, we prefer to use the notation $\hat\CL_{SU(N)_1}[B]$ as opposed to simply using  $\CL_{SU(N)_1}[B]$. We would also like to point out that the two sides of the duality have a $\Z_N$ one-form symmetry.  In the left hand side it acts on $SU(N)_1$ and in the right hand side it acts on $U(1)_{-N}$.

\subsec{Gauging the previous example $U(N)_{1,\pm N+1} \longleftrightarrow  U(1)_{- N\mp 1}$}

We start with \simpledua\ and turn $B$ into a dynamical field, which we denote by $e$ and couple it to a $U(1)$ background field $C$ in the two sides.  This is the operation $S$ in \twoope\ \WittenYA.  The left hand side becomes
\eqn\Udo{
{1\over 4\pi} \Tr \left[bdb-{2i\over 3}b^3\right]
+{1\over 2\pi} (c+A)d(\Tr b+e) +{1\over 4\pi} ede-{1\over 2\pi} edC +\CL_0[x,y;A] ~.}
Integrating out $c$ sets $e=-\Tr b$ and we have
\eqn\Udos{\eqalign{
&{1\over 4\pi} \Tr \left[bdb-{2i\over 3}b^3\right] +{1\over 4\pi}(\Tr b) d(\Tr b)
+{1\over 2\pi} (\Tr b)dC  +\CL_0[x,y;A] \cr &\qquad =\CL_{U(N)_{1,N+1}}[b;C]+\CL_0[x,y;A]~.}}

Repeating these operations in the right hand side of  \simpledua\ we find
\eqn\simpleduag{\eqalign{
&-{N\over 4\pi} udu +{1\over 2\pi }ud(e-NA) -{1\over 2\pi} edC +{1\over 2\pi} edA +{1\over 4\pi} ede-N{\cal I}[g,A]=\cr
&\CL_{U(1)_{-N-1}}[u;C-(N+1)A]+\CL_{U(1)_1}[e+u; A-C]-N{\cal I}[g,A] ~.}}
This establishes the duality
\eqn\thirddu{\eqalign{\hat \CL_{U(N)_{1,N+1}}[C] \quad \longleftrightarrow \quad  \hat \CL_{U(1)_{-N-1}}&[C-(N+1)A]-{1\over 4\pi} CdC+{1\over 2\pi} Cd A\cr
& -(N+1){\cal I}[g,A]  ~,}}
or more briefly $U(N)_{1,N+1}\longleftrightarrow U(1)_{-N-1}$.

In the special case $N=1$ we find the $U(1)_2\longleftrightarrow U(1)_{-2}$ duality of section 3.1, so this is a generalization of that duality.
All the comments we mentioned after equations \dualtola\ about that duality are applicable here.

We can repeat the discussion in this subsection after adding $-{1\over 2\pi} BdB - {1\over 2\pi }BdC$ to \simpledua. In terms of \twoope\ this is the transformation $ST^{-2}$. Then, instead of \Udos\ the left hand side becomes
\eqn\Udosr{\eqalign{
&{1\over 4\pi} \Tr \left[bdb-{2i\over 3}b^3\right] -{1\over 4\pi}(\Tr b) d(\Tr b)
+{1\over 2\pi} (\Tr b)dC  +\CL_0[x,y;A] \cr &\qquad =\CL_{U(N)_{1,-N+1}}[b;C]+\CL_0[x,y;A]~.}}
And instead of \simpleduag\ the right hand side becomes
\eqn\simpleduagr{
\CL_{U(1)_{-N+1}}[u;-C-(N-1)A]+\CL_{U(1)_{-1}}[e-u; A-C]-N{\cal I}[g,A] ~.}
We derive the duality
\eqn\thirddur{\eqalign{\hat \CL_{U(N)_{1,-N+1}}[C] \quad \longleftrightarrow \quad  \hat \CL_{U(1)_{-N+1}}&[-C-(N-1)A]+{1\over 4\pi} CdC-{1\over 2\pi} Cd A\cr
& -(N-1){\cal I}[g,A]  ~,}}
or more briefly $U(N)_{1,-N+1}\longleftrightarrow U(1)_{-N+1}$.

\newsec{Level/rank duality}

\subsec{The basic non-Abelian level/rank duality}

In order to avoid cluttering the equations we take $N,K$ to be positive integers.

The $SU(N)_K$ WZW model is known to be equivalent to the GKO coset \GoddardVK
\eqn\inputdualCFT{
SU(N)_K \quad \longleftrightarrow\  \quad {SU(NK)_1\over SU(K)_N}~.
}
It was rigorously proven and further discussed in \refs{\XuNXA,\Rehren,\PaulyAMA,\VOstrik}.
Using the rules of \MooreYH\ for writing coset theories in Chern-Simons theory we can write  this duality as
\eqn\inputdualCS{
SU(N)_K \quad  \longleftrightarrow\quad   {SU(NK)_1 \times SU(K)_{-N}\over \Z_K}  ~.
}
The Lagrangian for the left hand side is simply $\CL_{SU(N)_K}[B]$.  The Lagrangian for the right hand side can be written as an $SU(NK)_1\times SU(K)_{-N}$ Lagrangian and then mod out the gauge group by $\Z_K$ by allowing additional bundles.
Instead, we would like to follow our previous discussion and use $U(NK)\times U(K)$ gauge fields $b$ and $u$.

We write
\eqn\inputdualCSLh{\eqalign{
{1\over 4\pi}\Tr \left[bdb -{2i\over 3} b^3 \right]&-{N\over 4\pi}\Tr \left[udu -{2i\over 3} u^3 \right]+{1\over 4\pi} \Big[(\Tr b) - N(\Tr u)\Big] d\Big[(\Tr b) - N(\Tr u)\Big]
\cr
&+{1\over 2\pi} cd\Big[(\Tr b) - N(\Tr u)\Big] +{1\over 2\pi} (Kc+\Tr u)dB ~,
}}
where $c$ is a $U(1)$ gauge field.  The first two terms are as in the Lagrangians for $U(NK)_1$ and $U(K)_{-N}$.  The third term, which we will further motivate below, is such the $U(NK)\times U(K)$ gauge theory is not a spin theory.  First, let us ignore the term with $B$.  The term with $c$ correlates the gauge fields of $U(1)\subset U(NK) $ and of $U(1)\subset U(K) $, thus removing one degree of freedom. Also, it states that the flux of $\Tr b$ is a multiple of $N$, which is needed for the identification of this Lagrangian with the right hand side of \inputdualCS.  Next, we consider the term with $B$. Now, nontrivial $B$ can lead to a shift of the fluxes of $b$ (as in the $SU(N)$ theory).

The Lagrangian \inputdualCSLh\ was designed such that it is invariant under the one-form gauge symmetry \refs{\KapustinGUA,\GaiottoKFA} $b\to b+\xi\unit$, $u\to u + \xi\unit$, $c\to c-\xi$ with $\xi$ a $U(1)$ gauge field.  (The third term in \inputdualCSLh\ and the coupling to $B$ were motivated by this invariance.) This one-form gauge symmetry removes another degree of freedom and establishes that this Lagrangian describes the right hand side of \inputdualCS.

The fact that the one-form gauge symmetry removes a degree of freedom can be made manifest by shifting $b\to b-c\unit$ and $u\to u-c\unit$ in \inputdualCSLh\ (or equivalently, fixing the gauge $c=0$) to find
\eqn\inputdualCSL{\eqalign{
\CL_{SU(NK)_1\over SU(K)_N}[B]\equiv &{1\over 4\pi}\Tr \left[ bd b -{2i\over 3}  b^3 \right] -{N\over 4\pi}\Tr \left[ ud  u -{2i\over 3}  u^3 \right]\cr
&+{1\over 4\pi} \Big[(\Tr  b) - N(\Tr  u)\Big] d\Big[(\Tr  b) - N(\Tr u)\Big]
+{1\over 2\pi} (\Tr  u)dB ~.
}}
As a check that this Lagrangian indeed describes $SU(NK)_1\over SU(K)_N$, we can write the gauge group in \inputdualCSL\ as $U(NK)\times U(K)= SU(NK)\times SU(K) \times U(1)\times U(1)/\Z_{NK}\times \Z_K$.  The two $U(1)$ factors in the numerator have a $k$-matrix
\eqn\kmatrix{k=\pmatrix{NK(1+NK) & -(NK)^2 \cr -(NK)^2 & NK(-1+NK)}~,}
whose determinant is $-(NK)^2$.  The fact that it is negative shows that these two $U(1)$ factors do not contribute to the framing anomaly.  Also, this value of the determinant shows that this Abelian sector has $(NK)^2$ representations.  Therefore, the number of representations of this $ SU(NK)_1\times SU(K)_{-N} \times U(1)\times U(1)/\Z_{NK}\times \Z_K$ theory is the same as the number of representations of $ SU(NK)_1\times SU(K)_{-N} / \Z_K$.

Next, we would like to determine the coefficient of a $BdB$ counterterm in the duality.
The currents coupled to $B$ in the $SU(N)_K$ and the $ SU(NK)_1\over SU(K)_N$  theories are
\eqn\currents{\eqalign{
&J_{SU(N)_K} ={1\over 2\pi} dc ={1\over 2\pi}\left({K\over N} +\epsilon_K\right)dB   \cr
&J_{SU(NK)_1\over SU(K)_N} ={1\over 2\pi} d\Tr u={1\over 2\pi}\left({K\over N}+K^2\right) dB ~, }}
where we used the equations of motion to express them in terms of the classical background $dB$.  Therefore, in order to match them across the duality we should add a Chern-Simons counterterm and write the duality as
\eqn\SUSUd{\CL_{SU(N)_K}[ B]\quad \longleftrightarrow\quad  \CL_{SU(NK)_1\over SU(K)_N}[B] +{\epsilon_K-K^2 \over 4\pi} BdB ~.}

We should emphasize that this duality is a duality between two non-spin-Chern-Simons theories.  Neither side of the duality depends on the choice of spin structure and the theories couple only to a background $U(1)$ gauge field $B$; there is no need for a coupling to a spin$_c$ connection $A$.

\subsec{$SU(N)_K  \longleftrightarrow  U(K)_{-N}$}

Next, we would like to add the effect of the gravitational line in the duality \inputdualCS,\SUSUd\ by adding $\CL_0[A]$ to the two sides.  In the left hand side we now have $\hat\CL_{SU(N)_K}[B]$ \SUh.  And in the right hand side we have the sum of the Lagrangian \inputdualCSL, the counterterm added in \SUSUd\ and $\CL_0[A]$
\eqn\SUUf{\eqalign{
 &{1\over 4\pi}\Tr \left[ bd b -{2i\over 3}  b^3 \right] -{N\over 4\pi}\Tr \left[ ud  u -{2i\over 3}  u^3 \right]+{1\over 4\pi} \Big[(\Tr  b) - N(\Tr  u)\Big] d\Big[(\Tr  b) - N(\Tr u)\Big]
\cr
&+{1\over 4\pi}xdx +{1\over 2\pi}xd(y+A)+{1\over 2\pi} (\Tr  u)dB+{\epsilon_K-K^2 \over 4\pi} BdB  ~.}}
Substituting here $b\to b-(x+y-KB)\unit$, $u\to u-(x+y-KB)\unit$, $x\to- x +\Tr b - N \Tr u +KB$, $y\to y+x-\Tr b +N \Tr u - KB$ it becomes
\eqn\SUUfb{\eqalign{
 &{1\over 4\pi}\Tr \left[ bd b -{2i\over 3}  b^3 \right] -{N\over 4\pi}\Tr \left[ ud  u -{2i\over 3}  u^3 \right]+{1\over 2\pi} (\Tr  u)d(B-NA) +{1\over 2\pi} (\Tr b)dA
\cr
&-{1\over 4\pi}xdx -{1\over 2\pi}xd(y+A)+{K\over 2\pi} BdA +{\epsilon_K \over 4\pi} BdB \cr
=& \CL_{U(NK)_1}[A] +\CL_{U(K)_{-N}}[B-NA]-\CL_0[A] +{K\over 2\pi} BdA+{\epsilon_K \over 4\pi} BdB ~.}}
We conclude that
\eqn\SUUd{\eqalign{
\hat \CL_{SU(N)_K}[B] \quad \longleftrightarrow\quad \hat \CL_{U(K)_{-N}}[B-NA] +{K\over 2\pi} BdA+{\epsilon_K\over 4\pi} BdB-NK{\cal I}[g,A]  ~.}}
As a check, note that for $K=1$ this agrees with \simpledua, which was checked explicitly by a change of variables.  Similarly, for $N=1$ the two sides of the duality are almost trivial theories, where all the dynamical fields can be integrated out.  This leads to ${K+\epsilon_K\over 4\pi} BdB $ in the two sides.

Finally, the orientation-reversal of \SUUd\ is
\eqn\SUUdr{\eqalign{
\hat \CL_{SU(N)_{-K}}[B] \quad \longleftrightarrow\quad \hat \CL_{U(K)_{N}}[-B+NA] -{K\over 2\pi} BdA-{\epsilon_K\over 4\pi} BdB+NK{\cal I}[g,A]  ~.}}
(Equivalently, this can be derived by adding to the two sides of \SUUd\ $-{K\over 2\pi} BdA-{\epsilon_K\over 4\pi} BdB-{1\over 2\pi} CdB$, promoting $B$ to a dynamical field $e$, and then integrating it out.  After $C\to -B$, $N\longleftrightarrow K$ and some rearrangement we end up with \SUUdr.)

%%%%%%%%%%%%%%%%%%%%%%%%%%%%%%%%%%%%%%%%%%%%%%%%%%%%%%%%%%%%%%%%%%%%%%%%%%%%%%%%%%%%%%%%%%%%%%

\subsec{$U(N)_{K,N+K}  \longleftrightarrow  U(K)_{-N,-(N+K)}$}

Given the duality \SUUd\ we can derive additional dualities by adding counterterms to the two sides and turning the background fields to dynamical ones, i.e.\ using the operations \twoope.
An example of that was mentioned after \SUUdr.

As another example, we add to the two sides of \SUUd\ the counterterm ${1-\epsilon_K \over 4\pi}BdB-{1\over 2\pi}Bd(C+(K+1)A)$ with another $U(1)$ background field $C$.  Then we turn $B$ into a dynamical field $e$.
 The left hand side of \SUUd\ becomes
\eqn\SUUdg{\eqalign{
&{K\over 4\pi} \Tr \left[bdb-{2i\over 3}b^3\right]
+{1\over 2\pi} (c+KA)d(\Tr b+e) +{1\over 4\pi} ede -{1\over 2\pi} ed(C +(K+1)A)\cr
&\qquad+\CL_0[x,y;A] ~.}}
Integrating out $c$ sets $e=-\Tr b$ and \SUUdg\ becomes
\eqn\SUUdg{\eqalign{
&{K\over 4\pi} \Tr \left[bdb-{2i\over 3}b^3\right] +{1\over 4\pi}(\Tr b) d(\Tr b)
 +{1\over 2\pi} (\Tr b)d(C+(K+1)A) +\CL_0[x,y;A] \cr
 &=  \CL_{U(N)_{K,N+K}}[b;C+(K+1)A]+\CL_0[x,y;A]~.}}
The right hand side of \SUUd\ becomes
\eqn\SUUdgdr{\eqalign{
&-{N\over 4\pi}\Tr \left[ ud  u -{2i\over 3}  u^3 \right]+{1\over 2\pi} (\Tr  u)d(e-NA) +{1\over 4\pi} ede-{1\over 2\pi }ed(C+A)-NK{\cal I}[g,A] \cr
&=\CL_{U(K)_{-N,-N-K}}[u;C+(1-N)A]+\CL_{U(1)_1}[e+\Tr u; -C -A]-NK{\cal I}[g,A] ~. }}
This establishes the duality
\eqn\SUUdogta{\eqalign{\hat \CL_{U(N)_{K,N+K}} [C+(K+1)A] \quad
\longleftrightarrow& \quad \hat \CL_{U(K)_{-N,-N-K}}[C+(1-N) A] \cr
&-{1\over 4\pi} CdC-{1\over 2\pi} Cd A-(NK+1){\cal I}[g,A]~.}}

As checks, note that it is consistent with \rulespinc\ and that for $K=1$ it coincides with \thirddu.  Also, under orientation-reversal combined with $C\to -C -2A$ the duality \SUUdogta\ is mapped to itself with $N\longleftrightarrow K$.

%%%%%%%%%%%%%%%%%%%%%%%%%%%%%%%%%%%%%%%%%%%
\subsec{$U(N)_{K,K-N}  \longleftrightarrow  U(K)_{-N,K-N}$}

Another duality is obtained by adding to the two sides of \SUUd\ the counterterm $-{\epsilon_K + 1 \over 4\pi}BdB-{1\over 2\pi}Bd(C+(K-1)A)$ with another background field $C$, and promoting $B$ to be a dynamical field $e$.
After integrating out $c$ and $e$ the resulting duality is
\eqn\UUrevn{\eqalign{
\hat{\CL}_{U(N)_{K,K-N}}[C+(K-1)A]
\quad\longleftrightarrow\quad
&\hat{\CL}_{U(K)_{-N,K-N}}[-C-(N-1)A]+{1\over 4\pi}CdC\cr
&\quad-{1\over 2\pi}CdA-(NK-1)\CI[g,A].
}
}

We can set $K=N$ in \UUrevn:
\eqn\psu{\eqalign{
\hat{\CL}_{U(N)_{N,0}}[C+(N-1)A]
\quad\longleftrightarrow\quad
&\hat{\CL}_{U(N)_{-N,0}}[-C-(N-1)A]+{1\over 4\pi}CdC\cr
&\quad-{1\over 2\pi}CdA-(N^2-1)\CI[g,A]~.
}
}
Since $U(N)_{N,0}=[SU(N)_N\times U(1)_0]/\Z_N$, the $U(1)$ field does not have a quadratic term in \psu.  Integrating over it constrains the background field $C+(N-1)A$.\foot{Unlike the $SU(N)$ theories discussed above, where the background field leads to $SU(N)/\Z_N$ bundles, here the partition function includes a sum over such twisted bundles and the background field determines their relative phases.} This leads to the duality:\foot{We could have started with \inputdualCS\ with $N=K$ and mod it out by its $\Z_N$ one-form symmetry to find
\eqn\psueqprec{
{SU(N)_N \over \Z_N}  \quad \longleftrightarrow\quad  {SU(N)_{-N}\over \Z_N} \times {SU(N^2)_1\over \Z_N}.
}
For even $N$ these theories are spin and for odd $N$ they are non-spin. In the latter case  $SU(N^2)_1/\Z_N$ has only one representation and its
framing anomaly is a multiple of 8.  (For example, $SU(9)_1/\Z_3 \longleftrightarrow E(8)_1$.)  Therefore, for odd $N$  \psueqprec\ can be written as ${SU(N)_N / \Z_N}  \quad \longleftrightarrow\quad  {SU(N)_{-N}/ \Z_N} $ without the need to add the transparent line.}
\eqn\psueq{
{SU(N)_N\over \Z_N} \quad  \longleftrightarrow \quad {SU(N)_{-N}\over \Z_N}.
}

Finally, as another check of all the dualities \SUUd,\SUUdr,\SUUdogta,\UUrevn, the coefficient of ${\cal I}[g,A]$ accounts for the difference in the framing anomaly between the Chern-Simons gauge theories in the two sides.

%%%%%%%%%%%%%%%%%%%%%%%%%%%%%%%%%%%%%%%%%%%%%%%%%%%%
\newsec{Boson/fermion duality in Chern-Simons-Matter theories}

Our goal in this section is to write a more complete and precise version of \Oferdg
\eqn\Oferdgs{\eqalign{
N_f\ {\rm scalars\ with\ } SU(N)_K \quad &\longleftrightarrow \quad N_f\ {\rm  fermions\ with }\ U(K)_{-N+{N_f\over 2}} \cr
N_f\ {\rm  scalars\ with\ } U(N)_K\quad &\longleftrightarrow \quad N_f\ {\rm  fermions\ with\ } SU(K)_{-N+{N_f\over 2}}\cr
N_f\ {\rm  scalars\ with\ }U(N)_{K,K\pm N}\quad &\longleftrightarrow \quad N_f\ {\rm  fermions\ with\ } U(K)_{-N+{N_f\over 2}, -N\mp K+{N_f\over 2}}
~. }}
Again, we point out that the third duality in \Oferdgs\ with the bottom sign is new, as it is based on the new level/rank duality \newduality.  In addition to \Oferdgs\ we have four similar dualities obtained from these by reversing the orientation.  These differ from \Oferdgs\ by a sign change of all the levels.

In this section we write the precise Lagrangians of these theories with the background fields and their necessary counterterms.  This will allow us to turn these background fields into dynamical fields and to derive some consistency conditions (like deriving some of these dualities from the others), and to find additional dualities.

Before we do that we should review some facts about the fermion determinant in three dimensions.
Consider a three-dimensional fermion coupled to a background spin$_c$ connection $\CA$ via the Dirac operator $\slashchar{D} _\CA$
\eqn\freef{i\bar \psi \slashchar{D} _{\CA}\psi ~.}
Following \refs{\AlvarezGaumeNF,\WittenABA} the fermion determinant is
\eqn\Ztd{Z(\CA)=|\det \slashchar{D} _\CA|e^{- {i\pi\over 2}\eta(\CA)}~,}
where the sign in the exponent depends on the regularization and we pick it to be negative.
If we give the fermion a positive mass and integrate it out, the low energy theory has no induced Chern-Simons term.  And if the mass is negative, the low energy theory includes $-{\cal I}[g,\CA] $.

In the literature it is common to refer to the fermion coupled to $\CA$ with this regularization as $U(1)_{-\half}$ and describe it as having a bare counterterm $-{\cal I}[g,\CA]/2$.  We will follow this terminology and will refer to this theory as $U(1)_{-\half}$.  But we will not add this improperly quantized Chern-Simons term.  In more detail, although
\eqn\etaCIr{e^{i\pi\eta(\CA)} = e^{i{\cal I}[g,\CA]}~,}
we can take the square root of the left had side as the meaningful and gauge invariant expression $e^{i\pi\eta(\CA)/2}$ (which appears in \Ztd), but we cannot write the square root of the right hand side as $e^{i{\cal I}[g,\CA]/2}$, which is not meaningful.

In the various theories the gauge group is either $SU(L)$ or $U(L)$ with $L$ being $N$ in the scalar theories and $K$ in the fermionic theories.  As in section 2.4, we are going to represent the $SU(L)$ theory in terms of a $U(L)$ theory with a constraint.  So our bosonic covariant derivative $D$ includes a $U(N)$ gauge field and the Dirac operator $\slashchar{D}$ includes $\CA=u+A\unit$ with $u$ a $U(K)$ gauge field and $A$ a background spin$_c$ connection.  Also, in the bosonic side of the duality we include a $|\phi|^4$ term to denote the fact that the theory is at the corresponding Wilson-Fisher fixed point.  We will elaborate on this potential momentarily.

We claim that the dualities \Oferdgs\ are
\eqn\CSdualsmatter{\eqalign{
&|D\phi|^2-|\phi|^4+\hat \CL_{SU(N)_K}[B]\quad  \longleftrightarrow\quad i\bar\psi\slashchar{D}\psi+\hat\CL_{U(K) _{N_f -N}}[B-(N-N_f)A]\cr
&\qquad\qquad\qquad\qquad\qquad\qquad\qquad\quad
+{K\over 2\pi}B dA+{\epsilon_K\over 4\pi}BdB-(N-N_f)K{\cal I}[g,A]
\cr
&|D\phi|^2-|\phi|^4+ \hat \CL_{U(N)_{K}}[B+KA] \quad  \longleftrightarrow\quad
i\bar\psi\slashchar{D}\psi+ \hat\CL_{SU(K) _{N_f -N}}[-B]\cr
&\qquad\qquad\qquad\qquad\qquad\qquad\qquad\quad-{N-N_f\over 2\pi}B dA+{\epsilon_{N-N_f}\over 4\pi}BdB -(N-N_f)K{\cal I}[g,A]
\cr
&|D\phi|^2-|\phi|^4+ \hat \CL_{U(N)_{K,K\pm N}}[B+(K\pm 1)A ]\quad  \longleftrightarrow \quad
i\bar\psi\slashchar{D}\psi\cr
&\qquad\qquad\qquad\qquad\qquad\qquad\qquad\quad
+\hat\CL_{U(K) _{N_f -N,N_f -N \mp K}}[\pm B+(N_f-N+1 )A ]\cr
&\qquad\qquad\qquad\qquad\qquad\qquad\qquad\quad
-{1\over 2\pi}B dA \mp{1\over 4\pi}BdB -((N-N_f)K \pm 1){\cal I}[g,A]
~.
}}
and their orientation reversed ones are
\eqn\orientationr{\eqalign{
&|D\phi|^2-|\phi|^4+\hat{\CL}_{SU(N)_{-K}}[B] \quad\longleftrightarrow\quad
i\bar\psi\slashchar{D}\psi +\hat{\CL}_{U(K)_{N}}[-B+NA] \cr
&\qquad\qquad\qquad\qquad\qquad\qquad\qquad\quad
-{K\over 2\pi}BdA - {\epsilon_K\over 4\pi}BdB + NK\CI[g,A]
\cr
&|D\phi|^2-|\phi|^4+\hat{\CL}_{U(N)_{-K}}[-B-KA] \quad\longleftrightarrow\quad
i\bar\psi\slashchar{D}\psi +\hat{\CL}_{SU(K)_{N}}[-B] \cr
&\qquad\qquad\qquad\qquad\qquad\qquad\qquad\quad
+{N\over 2\pi}BdA - {\epsilon_{N}\over 4\pi}BdB + NK\CI[g,A]
\cr
&|D\phi|^2-|\phi|^4+\hat{\CL}_{U(N)_{-K,-K \mp N}}[-B-(K\pm 1)A] \quad\longleftrightarrow\quad
i\bar\psi\slashchar{D}\psi \cr
&\qquad\qquad\qquad\qquad\qquad\qquad\qquad\quad
+\hat{\CL}_{U(K)_{N,N\pm K}}[\mp B+(N- 1)A] \cr
&\qquad\qquad\qquad\qquad\qquad\qquad\qquad\quad
+{1\over 2\pi}BdA \pm {1\over 4\pi}BdB+\left(NK \pm 1\right)\CI[g,A]
~.
}
}

For $N_f=0$ there are no matter fields and the dualities \CSdualsmatter,\orientationr\ go over to the four level/rank dualities \SUUd,\SUUdr,\SUUdogta,\UUrevn\
\eqn\SUUds{\eqalign{
&\hat \CL_{SU(N)_K}[B] \quad \longleftrightarrow\quad \hat\CL_{U(K)_{-N}}[B-NA] +{K\over 2\pi} BdA+{\epsilon_K\over 4\pi} BdB-NK{\cal I}[g,A]  \cr
&\hat\CL_{U(N)_{K}}[B+KA]\quad \longleftrightarrow\quad \hat\CL_{SU(K)_{-N}}[-B] -{N\over 2\pi} BdA+{\epsilon_N\over 4\pi} BdB-NK{\cal I}[g,A]  \cr
&\hat\CL_{U(N)_{K,K\pm N}}[B+(K\pm 1)A] \quad
\longleftrightarrow \quad  \hat\CL_{U(K)_{-N,-N\mp K}}[\pm B+(1-N) A]\mp {1\over 4\pi} BdB\cr
& \qquad\qquad\qquad\qquad\qquad\qquad\qquad\qquad
-{1\over 2\pi} Bd A -(NK \pm 1){\cal I}[g,A]
~}}
and their orientation reversed versions.
In fact, \CSdualsmatter\ is obtained from \SUUds\ by substituting $N\to N-N_f$ in the fermionic theory and \orientationr\ is obtained by simply changing the signs of the levels.

For $N_f=1$ there is a single flavor of $N$ scalars $\phi^i$ (with $i=1,...,N$) and $K$ fermions $\psi^I$ (with $I=1,...,K) $ and the $|\phi|^4$ interaction is $|\sum_i\bar \phi_i\phi^i|^2$.  The dualities refer to (assumed) nontrivial IR theories, where the coefficient of the mass terms $\sum_I\bar \psi_I \psi^I$ in the fermionic side and $\sum_i\bar\phi_i\phi^i$ in the bosonic side are tuned to the fixed point.  We denote this fine-tuned value of the mass by zero.

For $N_f>1$ the scalars and the fermions also have a flavor index $\phi^{ia}$ and $\psi^{Ia}$ (with $a=1,...,N_f$).\foot{We thank O.~Aharony and S.~Minwalla for a useful discussion on the following points.}  Starting the bosonic theory at a free UV fixed point, the $SU(N_f)$-invariant scalar potential can be a linear combination of $\Tr M$, $\Tr M^2$, and $(\Tr M)^2$ with $M_a^b= \sum_i\bar\phi_{ai}\phi^{ib}$.  We assume that this theory has a nontrivial IR fixed point with $SU(N_f)$ global symmetry with a single $SU(N_f)$-invariant relevant operator $\sum_{Ia}\bar\psi_{aI}\psi^{Ia} \longleftrightarrow \Tr M$.  The fixed point is achieved when the coefficient of this relevant operator is fine-tuned.  Again, the duality refers to this IR theory.

In section 4 we related the dualities \SUUds\ by starting with one of them, adding counterterms and promoting background fields to dynamical fields.  Exactly the same manipulations can be performed in \CSdualsmatter.  In particular, assuming one of these dualities, say the first one, we can derive the others.

In the special case $N=N_f=K=1$ these dualities go over to the dualities studied in \SeibergGMD.  The bottom sign in the last duality in \CSdualsmatter\ is our new duality, which becomes in this case
\eqn\newdumone{|D\phi|^2-|\phi|^4+ \hat \CL_{U(1)_{0}}[B ]\quad  \longleftrightarrow \quad i\bar\psi\slashchar{D}\psi +\hat\CL_{U(1) _{ 1}}[- B+A] +{\cal I}[g,A-B]~.}
As a check, it can be derived from gauging $B$ in equation (2.10) in \SeibergGMD.  Note that the left hand side is manifestly orientation-reversal invariant but the right hand is not.  Its orientation-reversal invariance will be one of the fermion/fermion dualities in section 6.

Assuming any of these dualities with fixed values of the integers $(N_f, N, K)$ we can deform the theory by a mass term for one of the flavors to derive a duality with fewer flavors. We turn on a mass term $\sum_I \bar\psi_{N_fI}\psi^{IN_f}$ in the fermionic side and $M_{N_f}^{N_f}= \sum_i\bar\phi_{N_fi}\phi^{iN_f}$ in the bosonic side.  Unlike the similar supersymmetric dualities, here we still need to fine tune an $SU(N_f-1)$ invariant mass term in order to hit a nontrivial IR fixed point.  In the fermionic side it is $ \sum_{aI}\bar\psi_{aI}\psi^{Ia}$ and in the bosonic side it is $\Tr M= \sum_{ai} \bar\phi_{ai}\phi^{ia}$.

In the bosonic side for one sign of the mass square the scalar becomes massive and we flow to the theory with $(N_f-1, N, K)$.  For the other sign of the mass square the gauge group is Higgsed and we flow to the theory with $(N_f-1, N-1, K)$.
In the fermionic side of the duality these two flows correspond to different signs of the fermion mass.  For positive mass there is no induced Chern-Simons term and for negative mass the Chern-Simons terms are shifted by  $-\CL_{U(K)_1}[u;A] - K{\cal I}[g,A] $ (recall that the fermion couples to $u+A\unit$ with $u$ a $U(K)$ gauge field).  This has the effect of shifting the level of the existing Chern-Simons terms in the fermionic side.  The dualities \CSdualsmatter\ are consistent with these flows.

This discussion shows that the duality \Oferdgs\ might be problematic for $N_f>N$.  In that case we can turn on masses and flow to $N=0$.  Now the left hand side of the duality is an empty gapped theory, while the theory in the right hand side seems like it is non-trivial.

It is instructive to understand the physical interpretation of the currents that the background fields $B$ and $A$ couple to in \CSdualsmatter.  The $SU(L)$ theories are described in terms of $U(L)$ fields $u$ with a Lagrange multiplier that sets $\Tr u = -B$.  Therefore, in these theories $B$ couples to the baryon number current.  More precisely, in the fermionic $SU(K)$ theory (the right hand side of the second duality in \CSdualsmatter) the baryon current couples to $B+KA$, which is a background $U(1)$ field for even $K$ and a spin$_c$ connection for odd $K$.  Indeed, the baryons in this theory are bosons for even $K$ and they are fermions for odd $K$.  In the various $U(L)$ theories in \CSdualsmatter\ there is no baryon number symmetry and $B$ couples to the monopole charge.  Depending on the levels of the $U(L)$ Chern-Simons theory it is or is not a spin theory.  (Recall that even when it is a spin theory, because of the coupling to $A$ it is independent of the spin structure.) When it is a spin theory, its monopole operator is a fermion and it couples to a background spin$_c$ connection, and when it is not spin the monopole operator couples to a background $U(1)$ field.  It is straightforward to check that the levels in \CSdualsmatter\ and the couplings to $A$ and $B$ are such that the monopole and baryon operators are mapped correctly between the two sides of the duality.  This was studied in  \RadicevicYLA\ and was one of the motivations of \AharonyMJS.

\medskip

%%%%%%%%%%%%%%%%%%%%%%%%%%%%%%%%%%%%%%%%%%%%%%%%%%%%%%%%%%%%%%%%%%%%%%%%%%%%%%%%%%%%%%%%%%%%%%%
%%%%%%%%%%%%%%%%%%%%%%%%%%%%%%%%%%%%%%%%%%%%%%%%%%%%%%%%%%%%%%%%%%%%%%%%%%%%%%%%%%%%%%%%%%%%%%%
\newsec{New boson/boson and fermion/fermion dualities}

Our goal in this section is to use the conjectured boson/fermion dualities \CSdualsmatter, \orientationr\ to derive new boson/boson and fermion/fermion dualities.
We will mainly limit ourselves to $N_f=1$ and only discuss an example of $N_f=2$ in section 6.3.

\subsec{Boson/boson dualities}

We set $K=1$ in the second duality of \CSdualsmatter\ and the second duality of \orientationr\ to find a triality
\eqn\bbi{\eqalign{
i\bar\psi\slashchar{D}_{A}\psi \quad  \longleftrightarrow\quad&
|D\phi|^2-|\phi|^4+ \hat \CL_{U(N)_1}[A]
+(N-1){\cal I}[g,A]\cr
\quad  \longleftrightarrow\quad&|D \phi|^2-|\phi|^4+ \hat \CL_{U(N)_{-1}}[-A]
-N{\cal I}[g,A] ~,}
}
where the $ \phi$ in the different Lagrangians are different fields.  We see that all the interacting bosonic theories on the right hand side are dual to a theory of a single free fermion. Surprisingly, although their Lagrangians depend on $N$, their physics is independent of $N$.  This amounts to an infinite set of dualities relating interacting bosonic theories.

The special case of $N=1$ leads to
\eqn\bbo{|D\phi|^2-|\phi|^4+ \hat \CL_{U(1)_1}[A]
 \quad  \longleftrightarrow\quad |D \phi|^2-|\phi|^4+ \hat \CL_{U(1)_{-1}}[-A]
-{\cal I}[g,A] ~,
}
which agrees with equation (2.6) of \SeibergGMD.

Given these dualities we can act on them with $T$ and $S$ of \twoope\ and find new dualities.  These are infinitely many purely bosonic theories, which are dual to a single fermion coupled to dynamical gauge fields, which in general have some Chern-Simons couplings.

For example, starting with the duality between the first bosonic theory in \bbi\ with general $N$ and the second bosonic theory in \bbi\ with $N=1$ we can add to the two sides of the duality the counterterms ${1\over 2\pi}CdA + {1\over 4\pi}CdC+(1-N)\CI[g,A]$ and then gauge $A$ by substituting $A\rightarrow A+u$ to find
\eqn\bbpv{\eqalign{
|D\phi|^2-|\phi|^4+ &\hat \CL_{SU(N)_1}[C]
 \quad  \longleftrightarrow\quad|D_b\phi|^2-|\phi|^4
-\left({1\over 4\pi}bdb+{1\over 2\pi}bdu+{N\over 4\pi}udu\right)
\cr
&-{1\over 2\pi}bdA+{1\over 2\pi}ud\left(C-NA\right)
+{1\over 4\pi}CdC+{1\over 2\pi}CdA-N\CI[g,A]
~.
}
}
In the special case $N=1$ we can integrate out $u$ and \bbpv\ becomes the well-known particle-vortex duality.

%%%%%%%%%%%%%%%%%%%%%%%%%%%%%%%%%%%%%%%%%%%%%%%%%%%%%%%%%%%%
\subsec{Fermion/fermion dualities}

Next we look for fermion/fermion dualities.
Set $N=1$ in the first duality of \CSdualsmatter\ and the first duality of \orientationr\ to find the triality
\eqn\ff{\eqalign{
|D_{-B}\phi|^2-|\phi|^4 +\CL_0[A]
&\quad  \longleftrightarrow\quad
i\bar\psi\slashchar{D}\psi+\hat\CL_{U(K) _{0}}[B]
-{K\over 4\pi} BdB + {K\over 2\pi} BdA\cr
\quad  \longleftrightarrow\quad&
i\bar\psi\slashchar{D}\psi+\hat\CL_{U(K) _{1}}[-B+A]
+{K\over 4\pi} BdB - {K\over 2\pi} BdA+K{\cal I}[g,A]
~.
} }
As in the previous subsection the fields $\psi$ in these different dual Lagrangians are different fields.  This leads to an infinite number of fermion/fermion dualities.  The duality between the two fermionic theories with $K=1$ in \ff\ agrees with equation (2.16) of \SeibergGMD\ (up to a trivial use of charge conjugation).

Again, as in the previous section we consider the duality between the first fermionic Lagrangian in \ff\ with $K=1$ and the second one with arbitrary $K$.  We add to the two sides the counterterms $-{K\over 4\pi}BdB+{1\over 2\pi}Bd(-C+KA)+{1\over 4\pi}CdC+{1\over 2\pi}CdA-K\CI[g,A]$ and gauge $B$ by denoting it by $u$ to find
\eqn\ffpv{\eqalign{
i\bar\psi\slashchar{D}\psi+\hat\CL_{SU(K) _{1}}[C]
\quad  \longleftrightarrow\quad
&i\bar\psi\slashchar{D}_{b+A}\psi
+{1\over 2\pi}bdu - {K+1\over 4\pi}udu +{1\over 2\pi}ud\left(-C+(K+1)A\right)\cr
&\qquad \qquad +{1\over 4\pi}CdC+{1\over 2\pi}CdA-K\CI[g,A]
~.
}
}
In the special case $K=1$ the duality \ffpv\ becomes after a trivial change of variables the fermion/fermion duality in equation (2.14) of \SeibergGMD\
\eqn\ffKo{\eqalign{
i\bar\psi\slashchar{D}_{A}\psi
\quad  \longleftrightarrow\quad
&i\bar\psi\slashchar{D}_{b+A}\psi
+{1\over 2\pi}bdu - {2\over 4\pi}udu +{2\over 2\pi}udA-\CI[g,A]
~.
}
}

\subsec{Self-duality of QED with two fermions}

In this section we are going to use the previous dualities to derive a fermion/fermion duality similar to that of \XuLXA.  Our derivation can be easily extended to lead to many other dualities.

We start by substituting $A\to A+X$ in \ffKo\ and $A\to A-X$ in its orientation reversed duality
\eqn\ffKogr{\eqalign{
i\bar\psi\slashchar{D}_{X+A}\psi&\quad  \longleftrightarrow\quad
i\bar\chi\slashchar{D}_a\chi + {1\over 2\pi}adu
-{2\over 4\pi} udu+{1\over 2\pi}ud(A+X)-\CI[g,A+X] \cr
i\bar\psi\slashchar{D}_{-X+A}\psi&\quad  \longleftrightarrow\quad
i\bar\chi\slashchar{D}_a\chi+{1\over 4\pi} ada - {1\over 2\pi}adu
+{2\over 4\pi} udu-{1\over 2\pi}ud(A-X)+2{\rm CS_{grav}}
~,
} }
where $a$, appearing as $b+A$ in \ffKo, is a dynamical spin$_c$ connection. (Note the parity anomaly.)  We denoted the fermions in the right hand side by $\chi$ to highlight that they are different than the fermions in the left hand side $\psi$.

The direct product of the theories \ffKogr\ is described by the sum of their Lagrangians
\eqn\AUngaugedx{\eqalign{
i\bar\psi_1\slashchar{D}_{A+X}\psi^1 +i\bar\psi_2\slashchar{D}_{A-X}\psi^2&\quad  \longleftrightarrow\quad
i\bar\chi_1\slashchar{D}_{a_1}\chi^1 +i\bar\chi_2\slashchar{D}_{a_2}\chi^2
+{1\over 4\pi}a_2da_2-{2\over 4\pi}u_1du_1
\cr
&\qquad\qquad  +{2\over 4\pi}u_2du_2+{1\over 2\pi}(a_1du_1-a_2du_2)
+{1\over 2\pi}(u_1-u_2)dA
\cr
&\qquad\qquad +{1\over 2\pi}(u_1+u_2)dX-{1\over 4\pi} (A+X)d(A+X)
~,
}
}
where the indices $1$ and $2$ label the two sectors.

Next we gauge $A$ by adding the counterterms $\frac{1}{2\pi}AdY -\frac{1}{4\pi}YdY+\CI[g,A]$ with $Y$ is a background $U(1)$ field (note that it is consistent with \rulespinc) and then promoting $A$ to a dynamical spin$_c$ connection $a$. This gauging couples the two sectors labeled by $1$ and $2$.  In the right hand side we integrate out most of the gauge fields and are left with a single dynamical spin$_c$ connection $\tilde a$
\eqn\AGaugedx{\eqalign{
i\bar\psi_1 &\slashchar{D}_{a+X}\psi^1
+i\bar\psi_2 \slashchar{D}_{a-X} \psi^2
+{1\over 4\pi}  a d a
+{1\over 2\pi}  a dY - {1\over 4\pi}YdY + 2{\rm CS_{grav}}
\cr
&\quad  \longleftrightarrow\quad
i\bar\chi_1\slashchar{D}_{\tilde a-Y}\chi^1
+i\bar\chi_2\slashchar{D}_{\tilde a+Y}\chi^2
+{1\over 4\pi} \tilde ad\tilde a
+{1\over 2\pi} \tilde a dX -{1\over 4\pi}XdX + 2{\rm CS_{grav}}
~.
}
}

The theory on the left hand side is QED coupled to two fermions $\psi^1$ and $\psi^2$.  The Chern-Simons term ${1\over 4\pi}ada$ arises from our conventions of the fermion determinant (see the discussion around  \freef-\etaCIr).  We will return to it below.  This theory
has a manifest $SU(2)^X\times U(1)^Y$ global symmetry. $X$ couples to the Cartan generator of $SU(2)^X$ and $Y$ couples to the global $U(1)^Y$ symmetry associated with the connection $a$.

The theory on the right hand side is also QED coupled to two fermions $\chi^1$ and $\chi^2$.  It also has a manifest $SU(2)^Y\times U(1)^X$ global symmetry.  But here $Y$ couples to the Cartan generator of $SU(2)^Y$ and $X$ couples to the global $U(1)^X$ associated with the connection $\tilde a$.

This means that the theory at the fixed point must have a global $SU(2)^X\times SU(2)^Y=Spin(4)$ symmetry.  Each dual description makes manifest only an $SU(2)\times U(1)$ subgroup of it, but these are different subgroups.  Equivalently, each of the Lagrangians in \AGaugedx\ has a classical $SU(2)\times U(1)$ symmetry, but the quantum theory has an enhanced quantum $SU(2)\times SU(2)=Spin(4)$ symmetry.

The way the classical global $U(1)$ symmetry is enhanced to a global $SU(2)$ quantum symmetry is common in supersymmetric dualities \refs{\IntriligatorEX,\AharonyBX}, and in fact, this particular enhancement occurs in the $\CN=4$ supersymmetric version of this theory \IntriligatorEX.  (See also \KachruRMA\ for a recent related discussion.)

Let us deform the theory on the left hand side of \AGaugedx\ by an $SU(2)$ invariant mass term $m\bar\psi_i\psi^i$ and find the low-energy theory
\eqn\massdt{\eqalign{
&{1\over 4\pi}  a d a
+{1\over 2\pi}  a dY - {1\over 4\pi}YdY + 2{\rm CS_{grav}} \quad\longleftrightarrow \quad - {2\over 4\pi}YdY \qquad\qquad\qquad\qquad\ \  {\rm for}\ m>0\cr
&-{1\over 4\pi}  a d a
+{1\over 2\pi}  a dY - {1\over 4\pi}YdY - {2\over 4\pi}XdX- 2{\rm CS_{grav}} \quad\longleftrightarrow \quad - {2\over 4\pi}XdX\ \qquad {\rm for}\ m<0 ~.}}
We find the same low-energy theories when we deform the theory on the right hand side of \AGaugedx\ by $-m\bar\chi_i\chi^i$.  This means that the operators are mapped as $\bar\psi_i\psi^i\longleftrightarrow - \bar\chi_i\chi^i$.

We claim that our theory has a global $SU(2)^X\times SU(2)^Y$ global symmetry and here we coupled background fields only to the Cartan generators.  As a check of our claims we note that the Lagrangians in \massdt\ are normalized such that we can extend them by adding the other background fields for $SU(2)^X\times SU(2)^Y$
 \eqn\massdts{\eqalign{
  U(1)^Y_{-2}&\quad \longrightarrow\quad  SU(2)^Y_{-1}\qquad {\rm for}\ m>0\cr
  U(1)^X_{-2}&\quad \longrightarrow\quad  SU(2)^X_{-1}\qquad {\rm for}\ m<0 ~.}}
(In this particular case if $X$ and $Y$ become dynamical fields, there is no need to add degrees of freedom when moving from the left to the right as these theories are dual to each other.)

Next, we would like to compare our result \AGaugedx\ with \XuLXA.  In addition to fitting this duality into a large web of dualities, i.e.\ deriving it by assuming other, well-motivated dualities, our analysis has three key differences relative to \XuLXA.
\item{1.} We have formulated the theory with spin$_c$ connections such that our formulation of the theory does not depend on a choice of spin structure.  In particular, our $a$ and $\tilde a$ are dynamical spin$_c$ connections and $X$ and $Y$ are background $U(1)$ fields, which in the quantum theory can be extended to $SU(2)$ fields.
\item{2.} We are more precise about the definition of the fermion determinant (see the discussion around \freef-\etaCIr).  To relate to \XuLXA\ we imprecisely describe each of our fermions that couples to some $\CA$ as having an improperly quantized Chern-Simons term $-{1\over 8\pi} \CA d\CA - \CSg$.  This means that in order to compare with \XuLXA\ we need to add in the left hand side of the duality $-{1\over 4\pi} a da -{1\over 4\pi} X dX- 2\CSg$ and to add in the right hand side of the duality $-{1\over 4\pi} \tilde a d\tilde a -{1\over 4\pi} Y dY- 2\CSg$.  After doing that \AGaugedx\ is almost the same as the duality in \XuLXA.  To make them identical we also need to add to the two sides of the duality the counterterms ${1\over 4\pi} (XdX+ YdY)$.
\item{3.} The counterterms ${1\over 4\pi} (XdX+ YdY)$ that we have to add to \AGaugedx\ in order to match with \XuLXA\ are incompatible with \rulespinc.  But what is worse is that if we add these counterterms and then deform the theory with fermion masses, the low-energy theory is $U(1)_{\pm 1}^X\times U(1)_{\mp 1}^Y$ (see equation (5) in \XuLXA), which unlike \massdt, \massdts\ cannot be extended to $SU(2)^X\times SU(2)^Y$.

Finally, we have also established the $Spin(4)$ global symmetry of the model, thus clarifying the relation to \refs{\SenthilJK, \XuLXA} (once some counterterms are added).  Note that we argued that the long distance behavior of the system is a non-trivial fixed point with a $Spin(4)$ global symmetry.  But that fixed point cannot be described by the strong coupling limit of a continuum Lagrangian based on a non-linear sigma model, whose target space is $S^3$.  Such a continuum theory is non-renormalizable.

\bigskip

\noindent {\bf Acknowledgments:}
We would like to thank O.~Aharony, Z.~Komargodski, S.~Minwalla, E.~Witten, and C.~Xu for useful discussions. The work of P.H. is supported by Physics Department of Princeton University.  NS was supported in part by DOE grant DE-SC0009988.  NS gratefully acknowledges support from the Simons Center for Geometry and Physics, Stony Brook University at which some of the research for this paper was performed.
Any opinions, findings, and conclusions or recommendations expressed in this material are those of the authors and do not necessarily reflect the views of the funding agencies.

%%%%%%%%%%%%%%%%%%%%
\appendix{A}{Often used equations}

In this appendix we collect some useful equations.

The following terms are valid counterterms on spin$_c$ three-manifold (with $B,C$ being $U(1)$ gauge fields and $A$ a spin$_c$ connection):
\eqn\rulespinca{\eqalign{
&{1\over 2\pi}  BdC \cr
&{1\over 4\pi}  BdB+{1\over 2\pi}  Bd A  \cr
& {\cal I}[g,A] \equiv{1\over 4\pi}  AdA + 2 \CSg \cr
&16 \CSg.
}}

$U(N)_1$ is dual to a classical field theory
 \eqn\UNoneia{\CL_{U(N)_1}[b;A] = {1\over 4\pi}\Tr [bdb-{2i\over 3}b^3]
+{1\over 2\pi} (\Tr b) dA\quad  \longleftrightarrow \quad - N{\cal I}[g,A] ~.}

The Lagrangians for $U(N)$ and $SU(N)$ are
\eqn\UKKCSaa{\eqalign{
&\CL_{U(N)_{K}} [b;C+KA]
\equiv {K\over 4\pi} \Tr \left[bdb-{2i\over 3}b^3\right]
+{1\over 2\pi} (\Tr b) d\left(C+KA\right)\cr
&\CL_{U(N)_{K,K\pm N}}[b;C+(K\pm 1)A]\equiv
{K\over 4\pi} \Tr \left[bdb-{2i\over 3}b^3\right]\pm{1 \over 4\pi} (\Tr b) d(\Tr b)\cr
& \qquad\qquad+{1\over 2\pi} (\Tr b)d\left( C+(K\pm 1)A\right)\cr
&\CL_{SU(N)_K}[b,c;B]
\equiv{K\over 4\pi} \Tr \left[bdb-{2i\over 3}b^3\right]
+{\epsilon_K \over 4\pi}(\Tr b)d(\Tr b)
+{1\over 2\pi} cd\left(\Tr b+B\right) \cr
&\epsilon_K=\cases{0& $K$ even \cr 1& $K$ odd and positive \cr -1 & $K$ odd and negative~.
}}}
And if we add to them the ``trivial'' theory, which includes the purely gravitational spin one-half line
\eqn\trivialtheory{\CL_0[x,y;A]\equiv{1\over 4\pi}xdx +{1\over 2\pi} xd(y+A)}
we have
\eqn\SUha{\eqalign{
&\hat \CL_{SU(N)_K}[B]
\equiv{ K\over 4\pi} \Tr \left[bdb-{2i\over 3}b^3\right]
+{1\over 2\pi} (c+KA)d( \Tr b +B)  +\CL_0[A]+{\epsilon_K\over 4\pi }BdB \cr
&\qquad\qquad\longleftrightarrow\quad \CL_{SU(N)_K}[B] +\CL_0[A] \cr
&\hat\CL_{U(N)_K}[B+KA]\equiv \CL_{U(N)_K}[B+KA] +\CL_0[A] \cr
&\hat\CL_{U(N)_{K,K\pm N}}[B+(K\pm 1)A]\equiv \CL_{U(N)_{K,K\pm N}}[B+(K\pm 1)A] +\CL_0[A]  ~.}}

\listrefs
\end